\documentclass[aps,prc,twocolumn,showpacs,showkeys,amsmath,amssymb]{revtex4}
\usepackage{graphicx,dcolumn,bm}

\newcommand{\fm}{\, \text{fm}}
\newcommand{\mev}{\, \text{MeV}}
\newcommand{\gcq}{\, \text{g}/\text{cm}^{3}}

\begin{document}
\title{Cluster Formation and The Virial Equation of State of 
Low-Density Nuclear Matter}
\author{C.J. Horowitz}
\email[E-mail:~]{horowit@indiana.edu}
\author{A. Schwenk}
\email[E-mail:~]{schwenk@indiana.edu}
\affiliation{Nuclear Theory Center and Department of Physics, 
Indiana University, Bloomington, IN 47408}


\begin{abstract}

We present the virial equation of state of low-density nuclear matter 
composed of neutrons, protons and alpha particles. The virial equation
of state is model-independent, and therefore sets a benchmark for
all nuclear equations of state at low densities. We calculate the second
virial coefficients for nucleon-nucleon, nucleon-alpha and alpha-alpha
interactions directly from the relevant binding energies and scattering 
phase shifts. The virial approach systematically takes into account 
contributions from bound nuclei and the resonant continuum, 
and consequently provides a framework to include strong-interaction 
corrections to nuclear statistical equilibrium 
models. The virial coefficients are used 
to make model-independent predictions for a variety of properties of nuclear 
matter over a range of densities, temperatures and compositions. Our 
results provide constraints on the physics of the neutrinosphere 
in supernovae. The resulting alpha particle concentration differs from all 
equations of state currently used in supernova simulations. Finally, the 
virial equation of state greatly improves our conceptual understanding of 
low-density nuclear matter.

\end{abstract}

\maketitle 

\section{Introduction}

What do we mean by nuclear matter at subsaturation density?  
This is an important conceptual question.  The binding energy of 
uniform nuclear matter has a minimum at saturation density 
$n_0 \approx 0.16 \, \text{fm}^{-3}$.  Below this density, 
the system can minimize its energy by forming clusters.  Therefore 
the study of low-density nuclear matter is a study of clusters.  
What are the properties of these clusters?  What role do they 
play in astrophysics?  What is the relationship between clusters 
formed in the thermodynamic limit for an infinite system and those 
formed in laboratory heavy-ion collisions?  In this paper, we 
answer these questions using the virial expansion for low-density 
nuclear matter.

Simulations of core-collapse supernovae~\cite{snsim1,snsim2}, giant stellar 
explosions, depend on properties of low-density nuclear matter near 
the neutrinosphere.  If one views a star in visible light, one sees 
the photosphere.  This is the surface of last scattering for photons. 
Supernovae emit $99\%$ of their energy in neutrinos, and if one views 
a supernova in neutrinos, one sees the neutrinosphere.  This is the 
surface of last scattering for neutrinos and occurs at a density 
where the neutrino mean-free path is comparable to the size of the system. 
Direct evidence for the temperature of the neutrinosphere yields 
roughly $T \approx 4 \mev$, from the small number of neutrinos 
detected in SN1987a~\cite{sn1987a1,sn1987a2}. The neutrinosphere density, 
$n \sim 10^{11} \gcq \sim 1/1000 \, n_0$, then follows from known 
cross sections for neutrinos with these energies.  
{\it In this paper, we use the virial expansion to make 
model-independent predictions for the equation of state and other 
properties of low-density nuclear matter near the neutrinosphere.}  
We will use these results in later works to calculate neutrino interactions.

Nuclear statistical equilibrium (NSE) models are commonly used in 
nuclear astrophysics~\cite{nse}.  These models describe low-density
nuclear matter as a system of non- or minimally-interacting nuclei with 
a distribution of neutron number $N$ and charge $Z$ determined 
from nuclear masses through statistical equilibrium.  
However, NSE models fail in an uncontrollable manner as the density 
increases and strong interactions between nuclei become important.  
One would like a model-independent way to include strong interactions 
and a well-defined criterion to determine the range of validity of 
NSE models.  In this paper we take a step towards this goal by 
considering a simple NSE model with neutrons ($n$), protons ($p$) 
and alpha ($\alpha$) particles.  The virial expansion is then used to 
systematically incorporate strong interactions using nucleon-nucleon
(NN), N$\alpha$ and $\alpha\alpha$ elastic scattering phase shifts.

Pure neutron matter at low density is a very interesting 
nearly-universal Fermi system. In the limit that the neutron-neutron
scattering length is large compared to the interparticle 
spacing and the range of the interaction is small, 
the properties of dilute Fermi gases are 
expected to be universal, independent of the details of the 
interaction.  In a separate paper~\cite{vEOSneut},
we have calculated the equation of state of low-density neutron
matter using the virial expansion, in order to assess quantitatively 
how close the system is to universal behavior at finite temperature.  
The equation of state of resonant and dilute Fermi gases has also been 
studied in laboratory experiments with trapped 
atoms~\cite{Thomas1,Thomas2,Salomon,Grimm}, and Ho 
{\it et al.} have used the virial expansion to describe Fermi
gases at high temperatures in the vicinity of Feshbach 
resonances~\cite{Ho1,Ho2}.

There are many theoretical approaches to low-density nuclear matter.  
Microscopic calculations start from NN and three-nucleon interactions
that reproduce NN scattering and selected few-nucleon data.
In addition to the conventional variational~\cite{FP,APR} 
and Brueckner~\cite{Brueckner1,Brueckner2} calculations, 
renormalization-group
methods coupled with effective field theory provide new insights
to nuclear forces and the possibility of a perturbative and thus
systematic approach to nuclear matter
with theoretical error estimates~\cite{Vlowk1,Vlowk2}. However, 
all wave functions that are typically used only include two-nucleon 
correlations and omit four-nucleon correlations that could be 
important to describe alpha particle formation at low densities.  
We note that the equation of
state calculation of Buchler and Coon~\cite{BC} is close in strategy to 
the virial equation of state, but it takes into account Pauli blocking
on the phase shifts and thus uses a model NN interaction.
Moreover, there are a variety of Skyrme-type~\cite{skyrme1,skyrme2,skyrme3} 
and relativistic mean-field~\cite{RMF1,RMF2,RMF3} 
parametrizations of the nuclear energy functional,
which are used to calculate ground-state energies and densities of
intermediate-mass and heavy nuclei.  However, it is not clear what 
form to take for the energy functional for a gas of nucleons 
at low densities, where the system forms clusters and the binding energy 
per nucleon is density-independent with $E/A \approx -16 \mev$.

There are also some phenomenological models that attempt to provide 
a reasonable behavior for the equation of state over a wide range of 
densities, temperatures and compositions.  The Lattimer-Swesty
equation of state~\cite{LS} is based on an extended liquid-drop model.  
This equation of state is almost universally used in modern supernova 
simulations~\cite{snsim1,snsim2}. In addition, the 
Shen-Toki-Oyamatsu-Sumiyoshi equation of state~\cite{Shen} is based 
on an approximate Thomas-Fermi calculation using a relativistic 
mean-field interaction.  However, it is not clear how 
reliable these models are, and there does not appear to be any way 
to systematically improve them.  

Recently, there are lattice simulations for neutron matter 
using effective field theory~\cite{Dean}, where the couplings
are fitted to reproduce NN scattering regularized on the same lattice.  
These calculations are promising, but are presently limited to low orders 
in the effective field theory expansion and to small lattices.  It is 
very useful to compare these lattice results to our virial equation of 
state at low densities.  In the virial approach it is simple to assess
how errors in the NN phase shifts, because of a low-order truncation
in the effective field theory, impact the equation of state.  The virial
equation of state thus provides a valuable check for the lattice results.

The virial expansion is a very general and old method for determining 
the equation of state of a dilute gas.  There are two major assumptions.  
First, that the system is in a gas phase and has undergone no phase 
transition with decreasing temperature or increasing density.  
Second, that the fugacity 
$z = e^{\mu/T}$ is small, so the partition function can be expanded 
in powers of $z$.  Here $\mu$ denotes the chemical potential and $T$ 
is the temperature.  The second virial coefficient $b_2$ describes 
the $z^2$ term in the fugacity
expansion of the partition function, while the 
third virial coefficient describes the $z^3$ term.  A general formula 
relating $b_2$ to the two-body elastic scattering phase shifts has been 
known for some time~\cite{b2}.  Little is known about the third virial 
coefficient~\cite{b3Pais,b3Bedaque}. The first application of the
virial expansion to hot matter formed in heavy-ion experiments was
by Pratt {\it et al.}~\cite{Pratt}. In addition, Venugopalan and 
Prakash~\cite{Venugo}, for example, use the 
virial expansion to study matter formed in relativistic heavy-ion 
collisions. In general, however, the latter work is interested in 
much higher temperatures, where thermal mesons are important.

A great advantage of the virial expansion 
is that it correctly includes both bound states and scattering 
resonances on an equal footing.  In contrast, NSE models only include 
bound state contributions.  Since the thermodynamics of the system is
continuous when a bound state moves into the resonant continuum, 
both weakly bound states and low-energy scattering resonances are 
equally important. This was explicitly demonstrated for
Fermi gases in the crossover region of a Feshbach resonance by Ho
and Mueller~\cite{Ho1}. The virial expansion in this paper includes 
contributions from the two-nucleon (deuteron) and four-nucleon 
(alpha particle) bound states and the relevant low-energy scattering 
resonances: the dominant ones being the $^1$S$_0$ two-nucleon resonance 
($^2$He or $^2$n), the P$_{3/2}$ N$\alpha$ resonance and the 
$\alpha\alpha$ $0^+$ resonance that is the ground state of $^8$Be.

This paper is organized as follows.  In Section~\ref{formalism}, we 
present our virial formalism.  Results are presented in 
Section~\ref{results} for the virial coefficients and
equilibrium properties, such as the composition, pressure, entropy, 
energy and symmetry energy.  We conclude and give an outlook in 
Section~\ref{conclusions}.   

\section{Formalism}
\label{formalism}

In this section, we introduce the virial expansion for a system of 
neutrons, protons, and alpha particles.  We first discuss our choice 
of $n$, $p$ and $\alpha$ particle degrees of freedom.  Then we present 
the virial equation of state expanded to second order in the fugacities 
and calculate the entropy and the energy.  Finally, we relate the 
second virial coefficients to the relevant scattering phase shifts.

\subsection{Nuclear Statistical Equilibrium}

There is a close relation between the virial expansion and NSE models.
The latter include a variety of bound states, while the virial 
expansion includes both bound states and scattering resonances. The
virial equation of state can therefore also be regarded as a systematic
extension of NSE models to take into account strong interactions in 
the resonant continuum.

The deuteron is automatically included in the virial expansion as a 
bound-state contribution to the second virial coefficient $b_2 \sim
e^{E_d/T}$ with experimental deuteron binding energy $E_d=2.22 \mev$.
However, the $\alpha$ particle has a large binding energy, $E_\alpha = 
28.3 \mev$, and this will only be included in the fourth virial
coefficient describing four-nucleon interactions. In general,
fourth-order contributions are expected to be small at low densities 
and high temperatures.  However, for lower temperatures $e^{E_\alpha/T}$ 
is very large.  This will lead to an unnaturally large fourth virial 
coefficient that will greatly reduce the radius of convergence of the 
nucleonic virial expansion.

Our solution to this problem is to rearrange the original nucleonic
virial expansion and include $\alpha$ particles explicitly. This allows 
us to retain the large $e^{E_\alpha/T}$ terms at lower order in the
virial expansion.  This procedure can be extended to explicitly include 
a variety of additional heavy nuclei that may be important at even 
higher densities.  We discuss the effects of heavier nuclei and of 
three-nucleon contributions in more detail below. In this first paper, 
for simplicity, we limit ourselves to $n$, $p$ and $\alpha$ particle 
degrees of freedom.  We note that three-nucleon bound state effects 
are expected to be small, because their binding energies are small compared to 
$E_\alpha$, and the system will thus favor the formation of alpha particles.

Therefore, we consider a low-density gas of $n$, $p$ and $\alpha$ particles.  
In chemical equilibrium, the neutron $\mu_n$, proton $\mu_p$ and 
alpha $\mu_\alpha$ chemical potentials satisfy
\begin{equation}
2\mu_p+2\mu_n=\mu_\alpha\,,
\end{equation}
so that the neutron $z_n=e^{\mu_n/T}$, proton $z_p=e^{\mu_p/T}$ 
and alpha $z_\alpha=e^{(\mu_\alpha+E_\alpha)/T}$ fugacities are related by
\begin{equation}
z_\alpha=z_p^2 \, z_n^2 \, e^{E_\alpha/T} \,.
\label{zalpha}
\end{equation}
As discussed above, we will consider $z_\alpha$ to be the same order 
as $z_n$ or $z_p$ and expand the partition function through second 
order in the fugacities $z_n$, $z_p$ and $z_\alpha$.  This is equivalent
to the virial expansion with only nucleon densities at high temperatures 
and reduces to the virial expansion of a pure $\alpha$ gas at low 
temperatures (assuming equal numbers of $n$ and $p$, $z_n=z_p$).
Moreover, it is equivalent to including only the bound-state
contribution in the fourth nucleon virial coefficient.

\subsection{Virial Equation of State}

We expand the grand-canonical partition function $\mathcal{Q}$ for a system 
of $n$, $p$ and $\alpha$ particles confined to a volume $V$ as
\begin{align}
\mathcal{Q}(z_n,z_p,z_\alpha,V,T) &= 1 + z_n Q_n + z_p Q_p 
+ z_\alpha Q_\alpha
\nonumber \\[1mm]
&+ z_n^2 Q_{nn} + z_p^2 Q_{pp} + z_\alpha^2 Q_{\alpha\alpha}
\nonumber \\[1mm]
&+ 2z_pz_n Q_{np}+ 2z_nz_\alpha Q_{\alpha n} \nonumber \\[1mm]
&+ 2z_pz_\alpha Q_{\alpha p} + \mathcal{O}(z_i^3) \,.
\end{align}
Here, $Q_i=Q_i(V,T)$ and  $Q_n$, $Q_p$ and $Q_\alpha$ are the partition 
functions for single particle $n$, $p$ or $\alpha$ systems
\begin{equation}
\frac{Q_n}{V} = \frac{Q_p}{V} = \frac{2}{\lambda^3} \quad
\text{and} \quad \frac{Q_\alpha}{V} = \frac{1}{\lambda_\alpha^3} \,,
\end{equation}
where $\lambda$ ($\lambda_\alpha$) denotes the nucleon ($\alpha$ particle)
thermal wavelength,
\begin{equation}
\lambda = (2\pi/mT)^{1/2} \quad \text{and} \quad \lambda_\alpha = 
(2\pi/m_\alpha T)^{1/2} \,.
\end{equation}
We use $m_n=m_p$ for the nucleon mass $m$ and $m_\alpha=4m$ for the alpha 
particle mass, thus $\lambda_\alpha = \lambda/2$.  Below, we will relate 
the two-particle partition functions $Q_{ij}=Q_{ij}(V,T)$ with $i,j=n,p,
\alpha$ to the second virial coefficients.

Next, we expand $\log \mathcal{Q}$ to second order in the fugacities
\begin{align}
\frac{\log \mathcal{Q}}{V} &= 2 \, \frac{z_n}{\lambda^3} + 2 \, 
\frac{z_p}{\lambda^3} + \frac{z_\alpha}{\lambda_\alpha^3} +
z_n^2 \, \frac{Q_{nn}-\frac{1}{2}Q_n^2}{V} \nonumber \\[1mm]
&+ z_p^2 \, \frac{Q_{pp}-\frac{1}{2}Q_p^2}{V} 
+ z_\alpha^2 \, \frac{Q_{\alpha\alpha}-\frac{1}{2}Q_\alpha^2}{V} 
\nonumber \\[1mm]
&+ 2 z_p z_n \frac{Q_{np}-\frac{1}{2}Q_nQ_p}{V} 
+ 2 z_p z_\alpha \frac{Q_{\alpha p}-\frac{1}{2}Q_\alpha Q_p}{V} 
\nonumber \\[1mm]
&+ 2 z_n z_\alpha \frac{Q_{\alpha n}-\frac{1}{2}Q_\alpha Q_n}{V} 
+ \mathcal{O}(z_i^3) \,.
\end{align}
We define the second neutron virial coefficient $b_n$ as
\begin{equation}
b_n = \frac{\lambda^3}{2V}(Q_{nn}-\frac{1}{2}Q_n^2) \approx 
\frac{\lambda^3}{2V}(Q_{pp}-\frac{1}{2}Q_p^2) \,,
\end{equation}
where the second approximation is due to neglecting the Coulomb 
interaction between protons and assumes charge-independent
nuclear interactions.  Likewise, the second alpha virial coefficient 
$b_\alpha$ is given by
\begin{equation}
b_\alpha=\frac{\lambda_\alpha^3}{V}(Q_{\alpha\alpha}-\frac{1}{2}Q_\alpha^2) \,,
\end{equation}
and the virial coefficients describing strong interactions between 
$pn$ and N$\alpha$ particles are
\begin{align}
b_{pn}&=\frac{\lambda^3}{2V}(Q_{np}-\frac{1}{2}Q_nQ_p) \,, \\[1mm]
b_{\alpha n}&=\frac{\lambda_\alpha^3}{V}(Q_{\alpha n}-\frac{1}{2}Q_\alpha Q_n)
\approx \frac{\lambda_\alpha^3}{V}(Q_{\alpha p}-\frac{1}{2}Q_\alpha Q_p) \,,
\end{align}
where again the second approximation neglects the Coulomb 
interaction and strong-interaction charge dependences.
Coulomb effects typically depend on the ratio of the Coulomb energy to the
thermal energy $\Gamma_i = Z_i^2 \alpha/a_i T$~\cite{Ichimaru}, where
$Z_i$ is the ion charge, $\alpha$ the fine structure constant, and
$a_i = (3/4 \pi n_i)^{1/3}$ with $n_i$ the ion density. For example, for
symmetric nuclear matter (assumed composed only of nucleons) at a density
$n=10^{12} \gcq$ and $T=4 \mev$ one has $\Gamma_p = 0.05$, and thus we
expect Coulomb effects to be small. However, this should be checked
explicitly in future work and Coulomb interactions may be more important
when we extend the virial approach to densities where heavy nuclei are
present.
Note that the choice of $\lambda_\alpha^3$ in the last 
equation (instead of $\lambda^3$) is our convention.  Below, we will 
calculate these second virial coefficients from microscopic NN, N$\alpha$  
and $\alpha\alpha$ elastic scattering phase shifts.

The pressure $P$ is given by the logarithm of the partition function,
which we truncate after second order in the fugacities. In terms of 
the second virial coefficients, the pressure can be written as
\begin{align}
\frac{P}{T}=\frac{\log \mathcal{Q}}{V}&=\frac{2}{\lambda^3}(z_n+z_p +
(z_n^2+z_p^2) b_n + 2 z_p z_n b_{pn}) \nonumber \\[1mm]
&+\frac{1}{\lambda_\alpha^3}(z_\alpha + z_\alpha^2 b_\alpha + 2 
z_\alpha (z_n+z_p) b_{\alpha n}) \,.
\label{p}
\end{align}
The $n$, $p$ and $\alpha$ densities follow from derivatives of $\log 
\mathcal{Q}$ with respect to the chemical potential or the fugacity,
\begin{equation}
n_i = z_i \biggl(\frac{\partial}{\partial z_i} \frac{\log \mathcal{Q}}{V}
\biggr)_{V,T} \,.
\end{equation}
The resulting $n$, $p$ and $\alpha$ particle densities are given by
\begin{align}
n_n &= \frac{2}{\lambda^3}(z_n+2z_n^2b_n+2z_pz_nb_{pn}+8z_\alpha z_n 
b_{\alpha n}) \,, \label{nn} \\[1mm]
n_p &= \frac{2}{\lambda^3}(z_p+2z_p^2b_n+2z_pz_nb_{pn}+8z_\alpha z_p 
b_{\alpha n}) \,, \label{np} \\[1mm]
n_\alpha &= \frac{1}{\lambda_\alpha^3}(z_\alpha+2z_\alpha^2b_\alpha+
2z_\alpha(z_n+z_p)b_{\alpha n}) \,, \label{nalpha}
\end{align}
where we have used $\lambda^3/\lambda_\alpha^3=8$.

These equations provide a parametric form for the virial equation of 
state that is thermodynamically consistent.  
For values of $z_n$ and $z_p$, the alpha fugacity $z_\alpha$ follows 
from chemical equilibrium $z_\alpha = z_p^2 \, z_n^2 \, e^{E_\alpha/T}$ 
and one can then calculate the pressure $P$ and composition, $n_n$, $n_p$ and 
$n_\alpha$, from Eqs.~(\ref{p},\ref{nn}-\ref{nalpha}). The total baryon 
density $n_b$ is given by
\begin{equation}
n_b=n_n+n_p+4n_\alpha \,,
\label{nb}
\end{equation}
and the proton fraction $Y_p$, the number of protons per baryon, by
\begin{equation}
Y_p=(n_p+2n_\alpha)/n_b \,.
\label{yp}
\end{equation}
The dependence of the baryon density and the 
proton fraction on $z_n$ and $z_p$ 
can be inverted to yield the virial equation of state in terms of
\begin{equation}
P = P\bigl(z_n(n_b,Y_p,T),z_p(n_b,Y_p,T),z_\alpha(z_n,z_p,T),T\bigr) \,.
\end{equation}
Note that these dependences are nonlinear and strongly temperature 
dependent, mainly due to the $e^{E_\alpha/T}$ term in the chemical 
equilibrium condition for $z_\alpha$. In practice, for a given proton
fraction, we use Eq.~(\ref{yp}) to determine the proton fugacity as
a function of the neutron one $z_n(z_p,Y_p,T)$, and generate the virial
equation of state in tabular form for a range of $z_n$. This maintains
the thermodynamic consistency of the virial equation of state.

\subsection{Virial Coefficients}

We relate the second virial coefficients to the relevant scattering 
phase shifts. This extends the standard results for spin-zero 
particles~\cite{b2} to include spin and isospin. The virial coefficient 
is related to the partition function of the two-particles system 
$\sum_{\text{states}} e^{-E_{2}/T}$, where the sum is over all 
two-particle states of energy $E_{2}$.  This sum can be converted to an
integral over relative momentum $k$ weighted by the density of states.  
The difference between the density of states of an interacting
and a free two-particle system can be expressed in terms of the derivative 
of the two-body phase shift $d\delta(k)/dk$~\cite{b2,huang}.  Finally,
if one integrates by parts, the virial coefficients can be calculated from
an integral over the scattering phase shifts summed over all partial waves
with two-body spin $S$, isospin $T$, orbital angular momentum $L$ and total 
angular momentum $J$ allowed by spin statistics.

The virial coefficient $b_n$ describes the interactions in pure neutron 
matter, and we have
\begin{equation}
b_n(T)=\frac{1}{2^{1/2}\,\pi T} \int_0^\infty dE \: e^{-E/2T} \, 
\delta^{\text{tot}}_n(E) - 2^{-5/2} \,.
\label{bn}
\end{equation}
Here, $-2^{-5/2}$ is the free Fermi gas contribution due to the Pauli 
principle and 
$\delta^{\text{tot}}_n(E)$ is the sum of the $T=1$ elastic scattering
phase shifts at laboratory energy $E$. This sum includes a degeneracy 
factor $(2J+1)$,
\begin{align}
\delta^{\text{tot}}_n(E) &= \sum_{S,L,J} (2J+1) \, 
\delta_{\,^{2S+1}\text{L}_J}(E)
\nonumber \\[1mm]
&=\delta_{^1\text{S}_0}+\delta_{^3\text{P}_0}+3\,\delta_{^3\text{P}_1}
+5\,\delta_{^3\text{P}_2}+5\,\delta_{^1\text{D}_2}+ \ldots
\label{deltantot}
\end{align}
Here and in the following we have neglected the effects of the mixing
parameters due to the tensor force. We expect that their contributions
to the second virial coefficients describing spin-averaged observables
vanish.

The $pn$ virial coefficient can be decomposed as
\begin{equation}
b_{pn}(T)=b_{\text{nuc}}(T)-b_n(T) \,,
\label{bpn}
\end{equation}
where $b_{\text{nuc}}$ is the second virial coefficient for symmetric 
nuclear matter.  With $b_{pn}$ and $b_n$ one can describe asymmetric 
matter with arbitrary proton fraction.  For $b_{\text{nuc}}$,
the contributions from the deuteron bound state and the scattering
continuum are given by
\begin{align}
b_{\text{nuc}}(T) &= \frac{3}{2^{1/2}}\bigl(e^{E_d/T}-1\bigr) -2^{-5/2}
\nonumber \\[1mm]
&+ \frac{1}{2^{3/2}\,\pi T} \int_0^\infty dE \: e^{-E/2T} \, 
\delta^{\text{tot}}_{\text{nuc}}(E) \,,
\label{bnuc}
\end{align}
where the term $-1$ in the deuteron contribution comes from the 
partial integration (the phase shift at zero energy being $\pi$ 
times the number of bound states) and the factor $3$ counts the 
spin-isospin degeneracy of the deuteron. The total phase shift 
$\delta^{\text{tot}}_{\text{nuc}}(E)$ for nuclear matter 
also receives contributions from $T=0$ states,
\begin{align}
\delta^{\text{tot}}_{\text{nuc}}(E) &= \sum_{S,L,J} (2J+1)(2T+1) \, 
\delta_{\,^{2S+1}\text{L}_J}(E)
\nonumber \\[1mm]
&=3\,\delta_{^1\text{S}_0}+3\,\delta_{^3\text{S}_1}
+3\,\delta_{^1\text{P}_1} \nonumber \\[1mm]
&+3\,\delta_{^3\text{P}_0}
+9\,\delta_{^3\text{P}_1}+15\,\delta_{^3\text{P}_2}
+15\,\delta_{^1\text{D}_2} \nonumber \\[1mm]
&+3\,\delta_{^3\text{D}_1}+5\,\delta_{^3\text{D}_2}
+7\,\delta_{^3\text{D}_3}+ \ldots
\label{deltanuctot}
\end{align}

For spin-zero alpha particles we can directly use the results 
of~\cite{b2,huang}, which give for the virial coefficient $b_\alpha$
\begin{equation}
b_\alpha(T)=\frac{2^{1/2}}{\pi T} \int_0^\infty dE \: e^{-E/2T} \,
\delta^{\text{tot}}_\alpha(E) + 2^{-5/2} \,,
\label{ba}
\end{equation}
where $+2^{-5/2}$ describes the contribution for a free Bose gas 
and the total phase shift $\delta^{\text{tot}}_\alpha(E)$ for elastic
$\alpha\alpha$ scattering is given by
\begin{align}
\delta^{\text{tot}}_\alpha(E) &= \sum_{L} (2L+1) \,
\delta_{\text{L}}(E)
\nonumber \\[1mm]
&=\delta_{\text{S}}+5\,\delta_{\text{D}}+9\,\delta_{\text{G}}
+13\,\delta_{\text{I}} + \ldots
\label{deltaaatot}
\end{align}
For the $\alpha\alpha$ virial coefficient and for the following
N$\alpha$ virial coefficient, we have not taken into account the effects of
inelasticities. For low temperatures, we expect their contributions 
to be small due to the tight binding of the alpha particle.

Finally, the N$\alpha$ virial coefficient is given by
\begin{equation}
b_{\alpha n}(T)=\biggl(\frac{5}{4}\biggr)^{1/2} \frac{1}{\pi T} 
\int_0^\infty dE \: e^{-4E/5T} \, \delta^{\text{tot}}_{\alpha n}(E) \,,
\label{ban}
\end{equation}
with total phase shift $\delta^{\text{tot}}_{\alpha n}(E)$ for N$\alpha$ 
scattering at nucleon laboratory energy $E$,
\begin{align}
\delta^{\text{tot}}_{\alpha n}(E) &= \sum_{L,J} (2J+1) \, 
\delta_{\text{L}_J}(E)
\nonumber \\[1mm]
&=2\,\delta_{\text{S}_{1/2}}+2\,\delta_{\text{P}_{1/2}}
+4\,\delta_{\text{P}_{3/2}}+4\,\delta_{\text{D}_{3/2}}
+6\,\delta_{\text{D}_{5/2}} \nonumber \\[1mm]
&+6\,\delta_{\text{F}_{5/2}}+8\,\delta_{\text{F}_{7/2}}+ \ldots
\label{deltaantot}
\end{align}
Once the four virial coefficients $b_n(T)$, $b_{pn}(T)$, $b_\alpha(T)$ 
and $b_{\alpha n}(T)$ have been calculated, the pressure is determined 
from Eq.~(\ref{p}) using fugacities $z_n$ and $z_p$ that reproduce 
the desired baryon density $n_b$ and proton fraction $Y_p$ using 
Eqs.~(\ref{zalpha},\ref{nn}-\ref{yp}).  

\subsection{Entropy and Energy}

The entropy $S$ and the energy $E$ can be calculated from the 
virial equation of state using thermodynamics~\cite{Ho1,huang}. 
The entropy density $s=S/V$ follows from
\begin{equation}
s=\biggl(\frac{\partial P}{\partial T}\biggr)_{\mu_i} \,.
\end{equation}
Here the temperature derivative is at constant $\mu_p$ and $\mu_n$,
and thus constant $\mu_\alpha$ due to chemical equilibrium. This
leads to
\begin{align}
s &= \frac{5P}{2T} - n_n\log z_n - n_p\log z_p - n_\alpha\log z_\alpha
\nonumber \\[1mm]
&+ \frac{2T}{\lambda^3} \bigl((z_n^2+z_p^2) b_n^\prime + 2 z_p z_n 
b_{pn}^\prime \bigr) \nonumber \\[1mm]
&+ \frac{T}{\lambda_\alpha^3} \bigl(z_\alpha^2 b_\alpha^\prime + 2 
z_\alpha (z_n+z_p) b_{\alpha n}^\prime \bigr) \,,
\label{s}
\end{align}
where $b_i^\prime$ denotes the temperature derivative of the second virial 
coefficients $b_i^\prime(T)=db_i(T)/dT$.  The energy density 
$\epsilon=E/V$ can be calculated from the entropy density by
\begin{align}
\epsilon &= Ts+\sum_{i=n,p,\alpha} n_i \mu_i - P \nonumber \\[1mm]
&=\frac{3}{2}P - n_\alpha E_\alpha + \frac{2T^2}{\lambda^3}
\bigl((z_n^2+z_p^2) b_n^\prime + 2 z_p z_n b_{pn}^\prime \bigr)
\nonumber \\[1mm]
&+ \frac{T^2}{\lambda_\alpha^3}\bigl(z_\alpha^2 b_\alpha^\prime 
+ 2 z_\alpha (z_n+z_p) b_{\alpha n}^\prime \bigr) \,.
\label{epsilon}
\end{align}
We emphasize that the $\alpha$ particle binding-energy contribution 
$n_\alpha E_\alpha$ is very important for the energy. Finally, the 
entropy per baryon $S/A$, energy per baryon $E/A$ and the free
energy per baryon $F/A$ are given by
\begin{equation}
\frac{S}{A} = \frac{s}{n_b} \: , \quad
\frac{E}{A} = \frac{\epsilon}{n_b} \quad \text{and} \quad
\frac{F}{A} = \frac{f}{n_b} \,,
\label{ea}
\end{equation}
with the free energy density $f=\epsilon - T s$.

\section{Results}
\label{results}

In this section, we present results for the second virial coefficients, 
the equation of state and the composition, as well as the entropy,
energy and symmetry energy.

\begin{figure}[b]
\begin{center}
\includegraphics[scale=0.36,clip=]{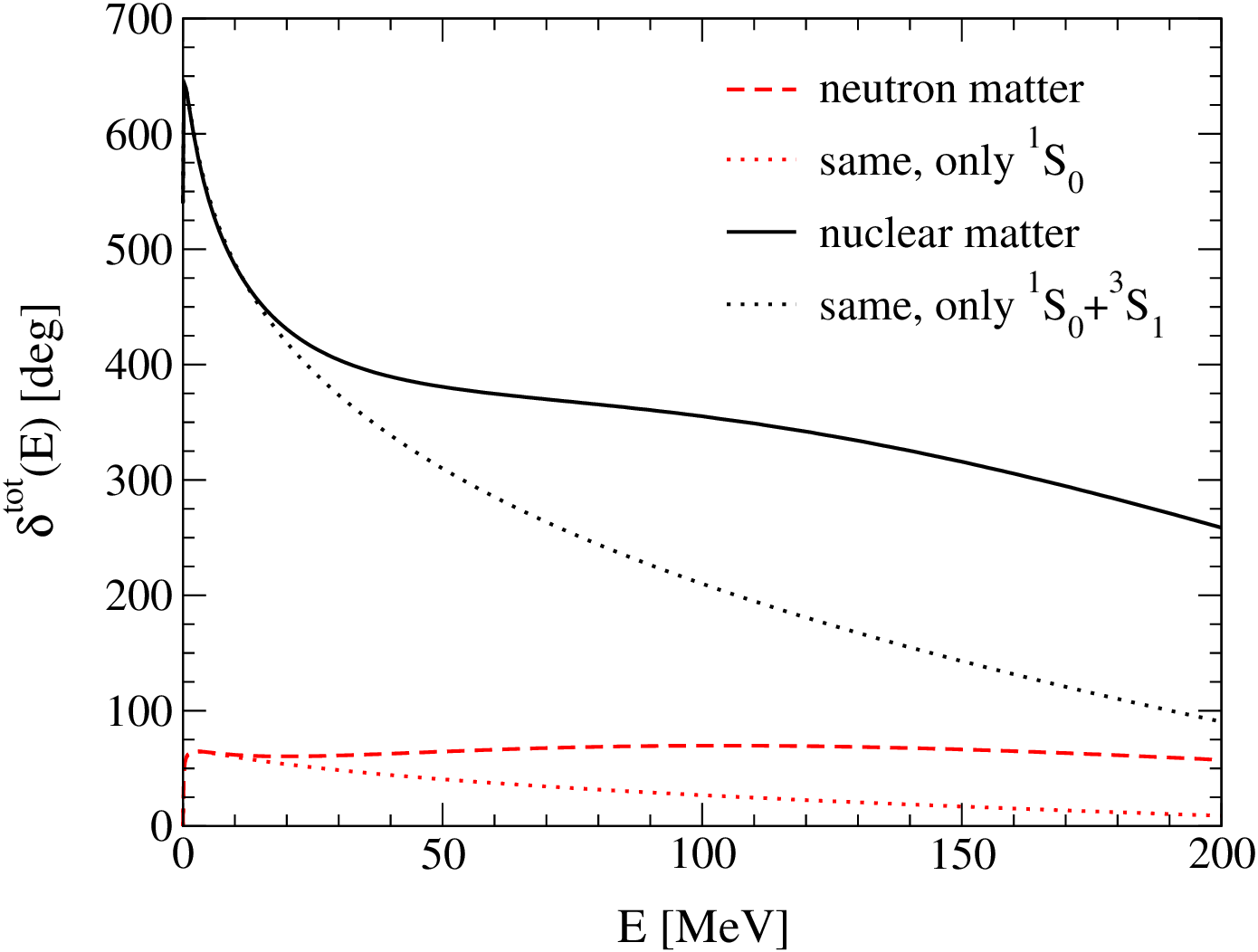}
\caption{(Color online) The total phase shift $\delta^{\text{tot}}_n(E)$ 
for neutron matter and $\delta^{\text{tot}}_{\text{nuc}}(E)$ for 
nuclear matter versus laboratory energy $E$. For reference, we
also show the contributions from only the S-wave phase shifts.}
\label{Fig1}
\end{center}
\end{figure}

\subsection{Virial Coefficients}

\begin{table*}[t]
\caption{NN virial coefficients for different temperatures. In
addition to the full results, we also give the virial coefficients 
calculated only from the large S-wave scattering lengths 
($a_{^1\text{S}_0} = -23.768 \fm$,
$a_{^3\text{S}_1} = 5.420 \fm$) and the deuteron binding energy.}
\label{Table1}
\begin{ruledtabular}
\begin{tabular}{c|ccc|ccc|ccc}
$T [\text{MeV}]$ & $b_n$ full & $b_n$ only $a_s$ & $T \, b_n^\prime$ full
& $b_{pn}$ full & $b_{pn}$ only $a_s, E_d$ & $T \, b_{pn}^\prime$ full
& $b_{\text{nuc}}$ full & $b_{\text{nuc}}$ only $a_s, E_d$ & $T \,
b_{\text{nuc}}^\prime$ full \\[0.5mm] \hline
1 & 0.288 & 0.357 & 0.032 & 19.4 & 19.5 & -43.8 & 19.7 & 19.8 
& -43.8 \\[0.5mm] 
2 & 0.303 & 0.400 & 0.012 & 6.10 & 6.24 & -7.39 & 6.40 & 6.64 
& -7.38 \\[0.5mm] 
3 & 0.306 & 0.421 & 0.005 & 4.01 & 4.19 & -3.54 & 4.31 & 4.61 
& -3.53 \\[0.5mm] 
4 & 0.307 & 0.434 & 0.002 & 3.19 & 3.40 & -2.30 & 3.50 & 3.83 
& -2.30 \\[0.5mm] 
5 & 0.308 & 0.443 & 0.002 & 2.74 & 2.98 & -1.73 & 3.05 & 3.43
& -1.72 \\[0.5mm] 
6 & 0.308 & 0.450 & 0.003 & 2.46 & 2.72 & -1.40 & 2.77 & 3.18
& -1.39 \\[0.5mm] 
7 & 0.308 & 0.456 & 0.004 & 2.26 & 2.54 & -1.18 & 2.57 & 3.00
& -1.18 \\[0.5mm] 
8 & 0.309 & 0.460 & 0.006 & 2.11 & 2.42 & -1.04 & 2.42 & 2.88
& -1.03 \\[0.5mm] 
9 & 0.310 & 0.464 & 0.008 & 2.00 & 2.32 & -0.93 & 2.31 & 2.79
& -0.92 \\[0.5mm] 
10 & 0.311 & 0.467 & 0.010 & 1.91 & 2.24 & -0.85 & 2.22 & 2.71
& -0.84 \\[0.5mm]
\end{tabular} 
\end{ruledtabular}
\end{table*}

We first calculate the NN virial coefficients. We take the NN phase 
shifts from the Nijmegen partial wave analysis~\cite{nnphases} and use 
$pn$ phase shifts for both $T=0$ and $T=1$ states. We thus neglect
the Coulomb interaction and strong-interaction charge dependences. 
In general, we expect that their effects on interaction energies
are small. All partial waves with $L \leqslant 2$ are included and 
we have checked that higher contributions are negligible for the 
temperatures of interest. The resulting total phase shifts for neutron 
and nuclear matter are shown in Fig.~\ref{Fig1}.

For neutron matter, we observe that $\delta^{\text{tot}}_n(E)$ is 
approximately energy-independent over a wide range.  The decrease 
of the $^1$S$_0$ phase shift with increasing energy is compensated 
by the contributions from higher angular momenta. As a result, 
the neutron virial coefficient $b_n$ will be approximately 
temperature-independent.  In contrast, $\delta^{\text{tot}}_{\text{nuc}}(E)$ 
decreases with increasing energy due to the decrease in the 
additional $^3$S$_1$ phase shift.  This behavior of the total phase, 
along with the strong temperature dependence of the deuteron 
contribution $e^{E_d/T}$, will lead to second virial coefficients
$b_{pn}$ and $b_{\text{nuc}}$ that decrease rapidly with temperature.

In Table~\ref{Table1}, we give our results for the NN virial 
coefficients $b_n(T)$, $b_{pn}(T)$ and $b_{\text{nuc}}(T)$, as
well as their derivatives $T b_i^\prime(T)$.
As expected, the NN virial coefficients are dominated by the
large S-wave scattering lengths and deuteron physics, but
effective range and higher partial wave contributions are
significant even for these relatively low temperatures. We also
emphasize that the effects of the low-energy $^1$S$_0$ resonance
encoded in $b_n$ become more important in neutron-rich matter.

\begin{figure}[t]
\begin{center}
\includegraphics[scale=0.36,clip=]{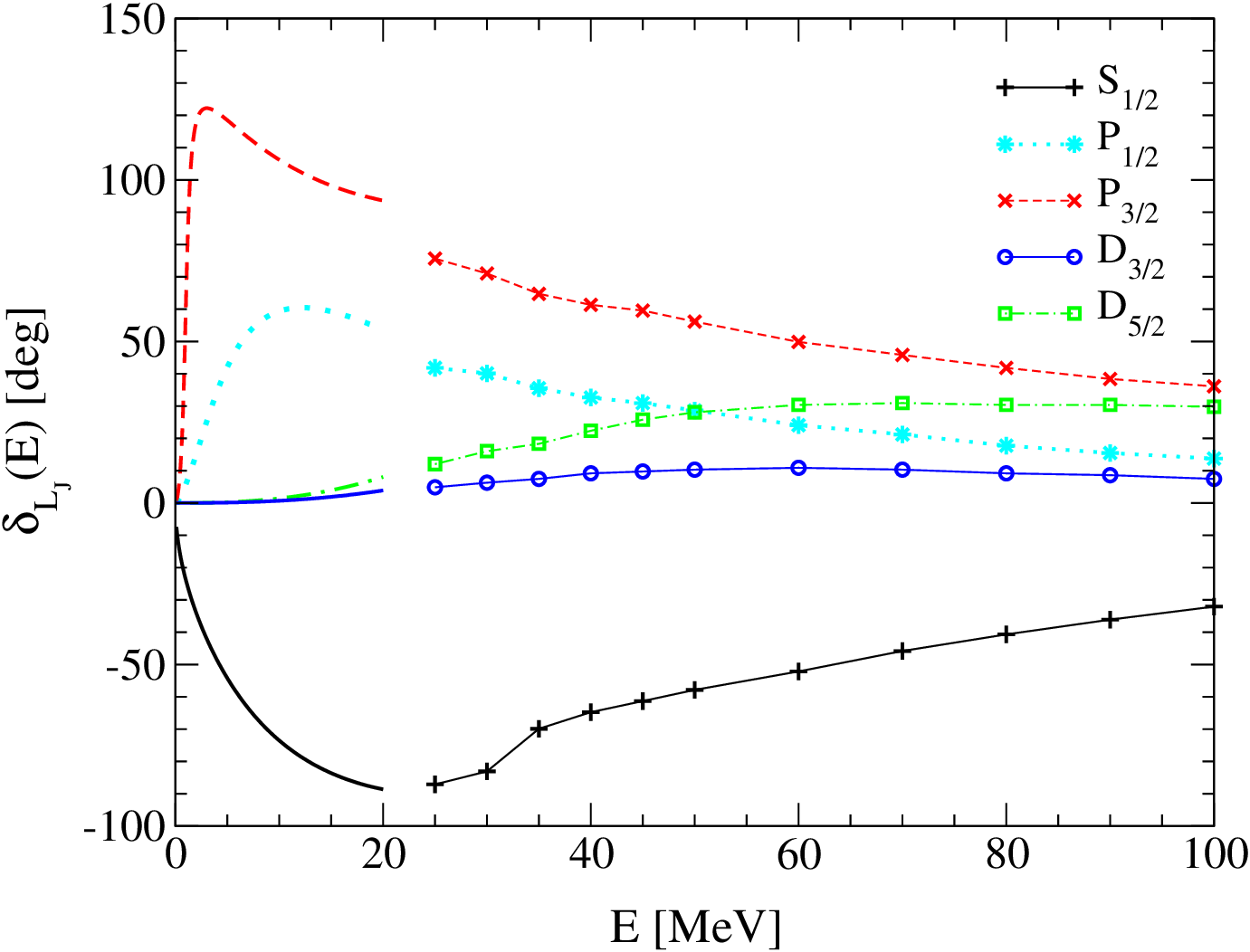}
\caption{(Color online) The phase shifts for elastic neutron-alpha
scattering $\delta_{\text{L}_J}(E)$ versus laboratory energy $E$. 
As discussed in the text, the
solid lines are from Arndt and Roper~\cite{alphanlow} and the symbols 
are from Amos and Karataglidis~\cite{alphanhigh}. For clarity, we
do not show the F-waves included in our results for $b_{\alpha n}$.}
\label{Fig2}
\end{center}
\end{figure}

Next, we calculate the N$\alpha$ virial coefficient. We neglect
Coulomb interactions and thus use neutron-alpha phase shifts
up to $L \leqslant 3$. For laboratory energies $E < 20 \mev$,
we take the simple phase shift fits of Arndt and 
Roper~\cite{alphanlow}, which are constrained by scattering
data over these energies. We extrapolate to higher energies
using microscopic phase shift predictions based on optical
model calculations by Amos and Karataglidis~\cite{alphanhigh}.
Note for these calculations we assume $|\delta_{\text{L}_J}|<\pi$.
Since we neglect the effects of inelasticities, we take the
real part of the predicted phase shifts of Amos and Karataglidis.
The differences between the real part and the absolute value
of the phase shifts are generally small over the relevant energies.
The neutron-alpha elastic scattering phase shifts are shown in
Fig.~\ref{Fig2}. The P$_{3/2}$ wave has a resonance at $E \approx
1 \mev$ with P$_{3/2}$ scattering length $a_{\text{P}_{3/2}} =
-62.95 \fm^{3}$~\cite{alphanERE}, and therefore will be the 
most important contribution to the N$\alpha$ virial coefficient.
The other partial waves are not resonant at low energies.
Our results for the N$\alpha$ virial coefficients are listed 
in Table~\ref{Table2}. We expect that more accurate high-energy 
phases will only slightly change our results for $b_{\alpha n}$ 
for temperatures $T > 5 \mev$.

\begin{table*}[t]
\caption{N$\alpha$ and $\alpha\alpha$ virial coefficients for 
different temperatures. In addition to the full results, we also 
give the virial coefficients calculated only from the dominant
low-energy resonances (P$_{3/2}$-wave for N$\alpha$, and S- and D-waves 
for $\alpha\alpha$).}
\label{Table2}
\begin{ruledtabular}
\begin{tabular}{c|ccc|ccc}
$T [\text{MeV}]$ & $b_{\alpha n}$ full & $b_{\alpha n}$ only P$_{3/2}$-wave &
$T \, b_{\alpha n}^\prime$  full & $b_\alpha$ full & $b_\alpha$ only S-
and D-waves & $T \, b_\alpha^\prime$ full \\[0.5mm] \hline
1 & 1.51 & 1.73 & 1.21 & 2.55 & 2.55 & 1.59 \\[0.5mm]
2 & 2.26 & 2.48 & 0.90 & 4.12 & 4.00 & 2.95 \\[0.5mm]
3 & 2.57 & 2.78 & 0.63 & 5.64 & 5.07 & 4.81 \\[0.5mm]
4 & 2.73 & 2.91 & 0.44 & 7.44 & 6.02 & 7.90 \\[0.5mm]
5 & 2.81 & 2.97 & 0.32 & 9.57 & 7.01 & 11.3 \\[0.5mm]
6 & 2.86 & 2.99 & 0.23 & 11.9 & 8.05 & 14.3 \\[0.5mm]
7 & 2.89 & 2.99 & 0.18 & 14.3 & 9.08 & 16.3 \\[0.5mm]
8 & 2.92 & 2.98 & 0.15 & 16.5 & 10.0 & 17.3 \\[0.5mm]
9 & 2.93 & 2.96 & 0.14 & 18.6 & 10.9 & 17.5 \\[0.5mm]
10 & 2.95 & 2.93 & 0.13 & 20.4 & 11.7 & 17.0 \\[0.5mm]
\end{tabular} 
\end{ruledtabular}
\end{table*}

\begin{figure}[t]
\begin{center}
\includegraphics[scale=0.36,clip=]{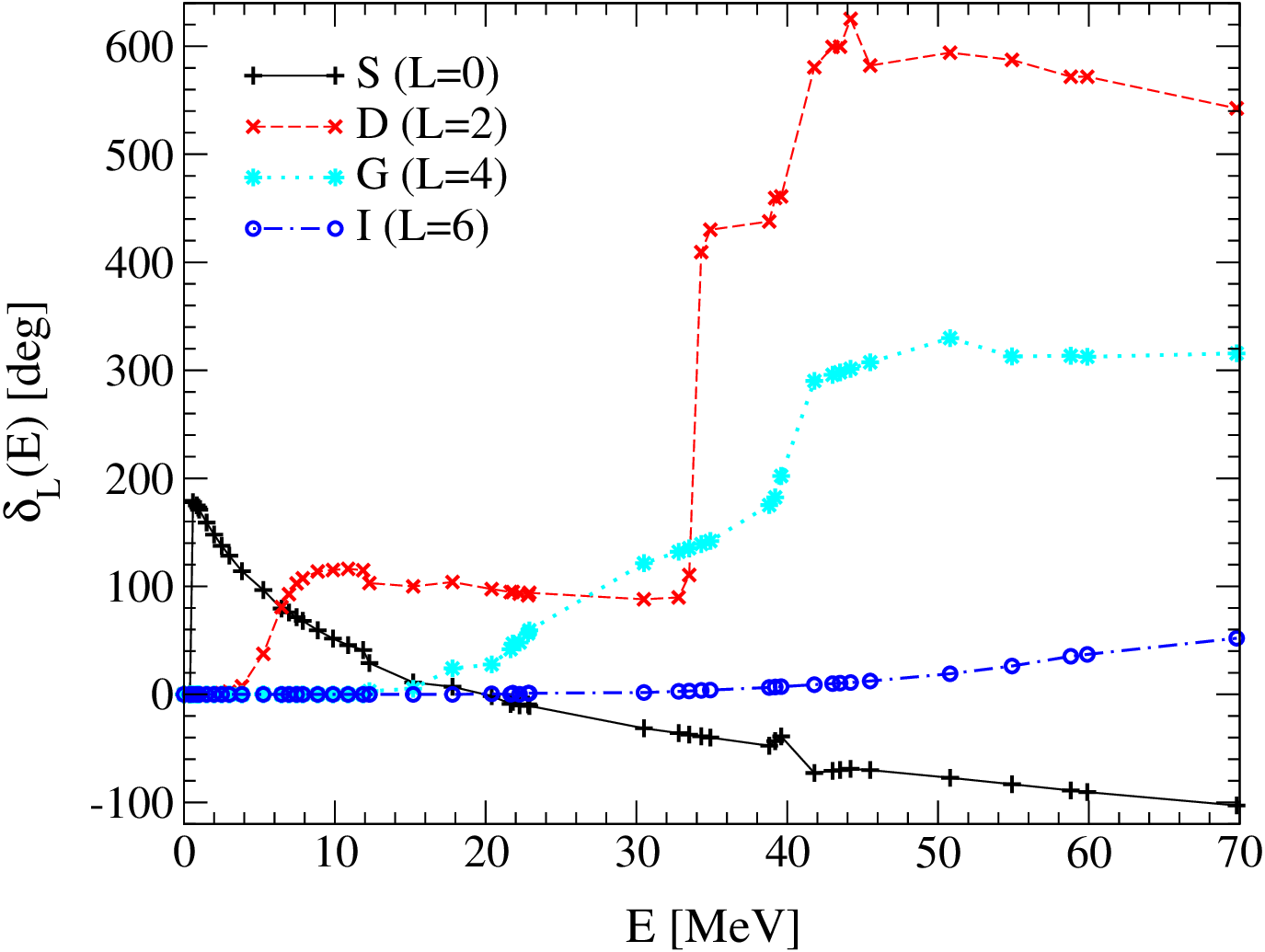}
\caption{(Color online) The phase shifts for elastic $\alpha\alpha$ 
scattering $\delta_{\text{L}}(E)$ versus laboratory energy $E$.  
As discussed in the text,
the phase shifts are taken from Afzal {\it et al.}~\cite{alphalow} 
and from Bacher {\it et al.}~\cite{alphahigh}.}
\label{Fig3}
\end{center}
\end{figure}

Finally, we show the phase shifts for elastic $\alpha\alpha$ scattering
in Fig.~\ref{Fig3}.  The low-energy phase shifts are taken from Afzal 
{\it et al.}~\cite{alphalow} and the phase shifts for energies between 
$30 \mev$ and $70 \mev$ are from Bacher {\it et al.}~\cite{alphahigh}.
The phase shifts display a pronounced $0^+$ resonance at very low 
energies that corresponds to the $^8$Be ground state.  This resonance 
is crucial for $^4$He burning in red giant stars.  Just below $35 \mev$ 
there are also two very close $2^+$ resonances (they are both at
$E \approx 34 \mev$ in Fig.~\ref{Fig3}) and in general the phase 
shifts become large at high energies. We include all $\alpha\alpha$ phases 
with $L \leqslant 6$ for the calculation of $b_\alpha$, where we make the
following approximations.  First, we neglect the Coulomb interaction.  
We also neglect inelastic channels.  Fortunately, the $\alpha$ particle 
is so tightly bound that important inelastic contributions do not arise 
until relatively high energies.  Finally, we truncate the integration 
for $b_\alpha$ at $70 \mev$ (the extent of the data).
Our results for the $\alpha\alpha$ virial 
coefficients are given in Table~\ref{Table2}. As for the NN virial 
coefficients, $b_{\alpha n}$ and $b_\alpha$ are dominated by the relevant 
low-energy resonances, but $b_{\alpha}$ receives important contributions
from higher angular momentum resonances for $T > 5 \mev$.

\vspace{0.2in}

In addition to the values for the NN, N$\alpha$ and $\alpha\alpha$ 
virial coefficients given in Tables~\ref{Table1} and~\ref{Table2}
for $T \leqslant 10 \mev$, we extend our results to higher temperatures
in Table~\ref{Table3}. We list these results in a separate table, as
our approximations, mainly the neglect of inelastic channels and the
truncation of the integration over the phase shifts at the extent of
the data, may be more severe.

\begin{table*}
\caption{NN, N$\alpha$ and $\alpha\alpha$ virial coefficients for 
higher temperatures. Note that inelasticities are neglected and
that the integration over the $\alpha\alpha$ elastic scattering phase 
shifts is truncated at $E \leqslant 70 \mev$.}
\label{Table3}
\begin{ruledtabular}
\begin{tabular}{c|cc|cc|cc|cc|cc}
$T [\text{MeV}]$ & $b_n$ full & $T \, b_n^\prime$ full
& $b_{pn}$ full & $T \, b_{pn}^\prime$ full
& $b_{\text{nuc}}$ full & $T \, b_{\text{nuc}}^\prime$ full 
& $b_{\alpha n}$ full & $T \, b_{\alpha n}^\prime$  full 
& $b_\alpha$ full & $T \, b_\alpha^\prime$ full \\[0.5mm] \hline
12 & 0.313 & 0.013 & 1.76 & -0.73 & 2.08 & 
-0.72 & 2.97 & 0.12 & 23.3 & 14.7 \\[0.5mm] 
14 & 0.315 & 0.014 & 1.66 & -0.65 & 1.97 & 
-0.64 & 2.98 & 0.10 & 25.4 & 11.7 \\[0.5mm] 
16 & 0.317 & 0.014 & 1.57 & -0.59 & 1.89 & 
-0.58 & 3.00 & 0.07 & 26.7 & 8.48
\\[0.5mm] 
18 & 0.319 & 0.013 & 1.51 & -0.55 & 1.82 & 
-0.54 & 3.00 & 0.02 & 27.5 & 5.44
\\[0.5mm] 
20 & 0.320 & 0.011 & 1.45 & -0.52 & 1.77 & 
-0.51 & 3.00 & -0.04 & 28.0 & 2.69
\end{tabular} 
\end{ruledtabular}
\end{table*}

\subsection{Composition}

\begin{figure}[t]
\begin{center}
\includegraphics[scale=0.36,clip=]{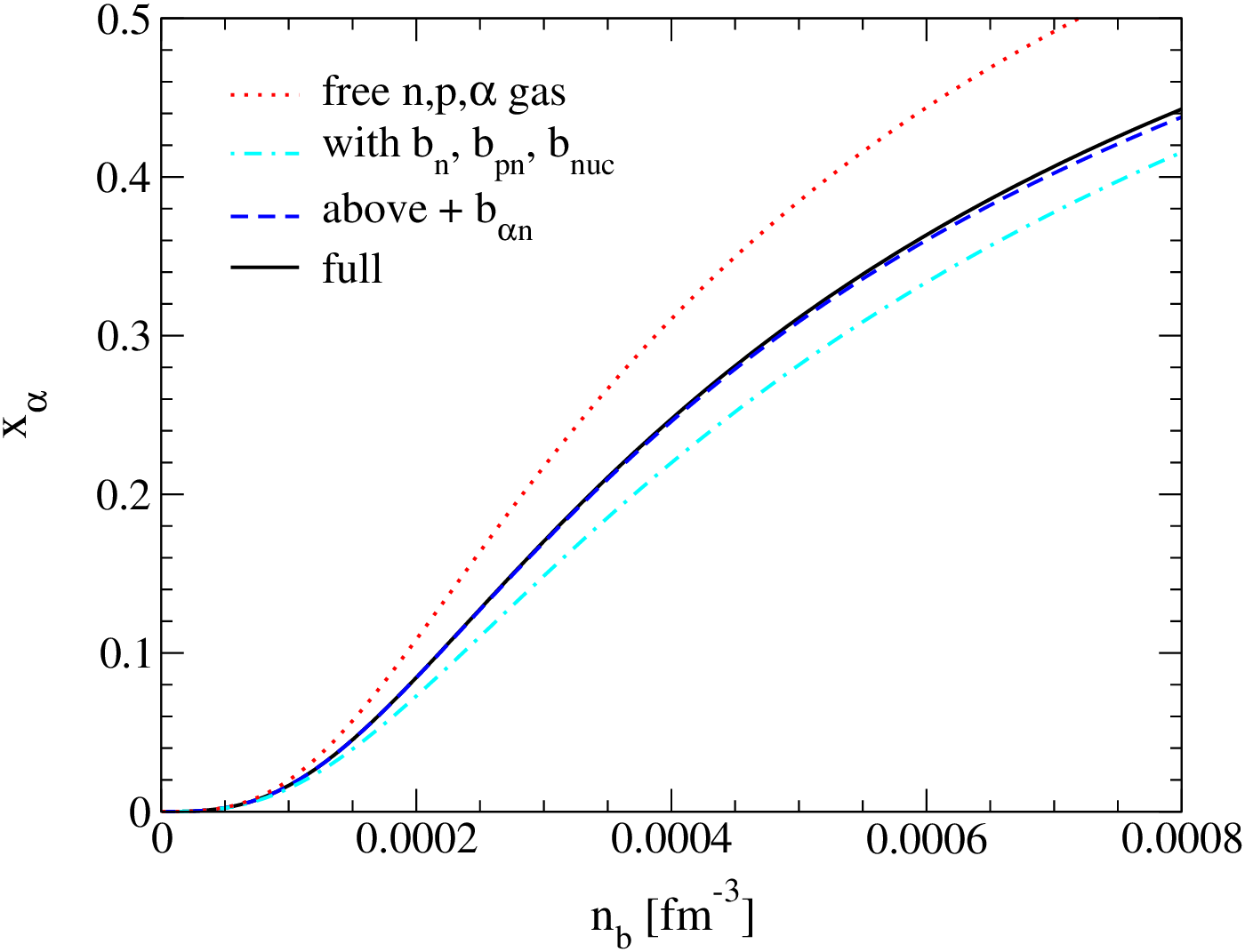}
\caption{(Color online) The $\alpha$ mass fraction $x_\alpha$ versus 
baryon density $n_b$ for $T = 4 \mev$, $Y_p = 1/2$ and various values 
of the virial coefficients, as explained in the text. The fugacities 
are small over this density range $z_n = z_p < 0.05$ and $z_\alpha < 0.004$.}
\label{Fig4}
\end{center}
\end{figure}

We now discuss the $\alpha$ particle concentration.  The $\alpha$ mass 
fraction is given by $x_\alpha=4n_\alpha/n_b$.  In Fig.~\ref{Fig4}, we 
show $x_\alpha$ for a temperature $T = 4 \mev$ and proton fraction 
$Y_p=1/2$ with various values of the virial coefficients.  
The largest $\alpha$ particle concentration results for 
a free $n$, $p$, $\alpha$ gas, where all $b_i = 0$. Including nucleonic
interactions $b_n, b_{pn}, b_{\text{nuc}} \neq 0$, but still keeping 
$b_\alpha, b_{\alpha n}=0$ reduces $x_\alpha$ significantly.  This is 
because the attractive nuclear interactions reduce the nucleon chemical 
potential, and this leads to a reduction of $x_\alpha$.  Including 
also $b_{\alpha n}\neq 0$ increases the $\alpha$ mass fraction,
but $x_\alpha$ is still smaller than its free value. The
latter effect is dominated by the low-energy N$\alpha$ resonance.
Finally, including $b_\alpha \neq 0$ only increases $x_\alpha$ by a very 
small amount. Therefore, in general, NN virial coefficients are expected 
to be most important, while the N$\alpha$ virial coefficient leads to
small changes, and $b_\alpha$ is least important.  This hierarchy is 
because typically $n_p+n_n > n_\alpha$.

The LS model excludes nucleons from the volume occupied by alpha particles. 
Although this is a prescription, it leads to an $\alpha$ mass fraction that
vanishes in the high-density limit~\cite{LPRL}. The effects of the excluded
volume are small at low densities. However, the excluded volume corresponds 
to a repulsive N$\alpha$ interaction that reduces the $\alpha$ particle 
concentration. In contrast, the neutron-alpha resonance and the attractive
total phase shift leads to an attractive N$\alpha$ intraction in the virial 
expansion. As a result, the $\alpha$ mass fraction increases, as shown in 
Fig.~\ref{Fig4}. Therefore, the excluded volume alone gives the wrong sign 
for N$\alpha$ interactions in the low-density limit, but we caution that 
the effects of the excluded volume and the N$\alpha$ virial coefficient may be 
small at very low densities.

\begin{figure}[t]
\begin{center}
\includegraphics[scale=0.36,clip=]{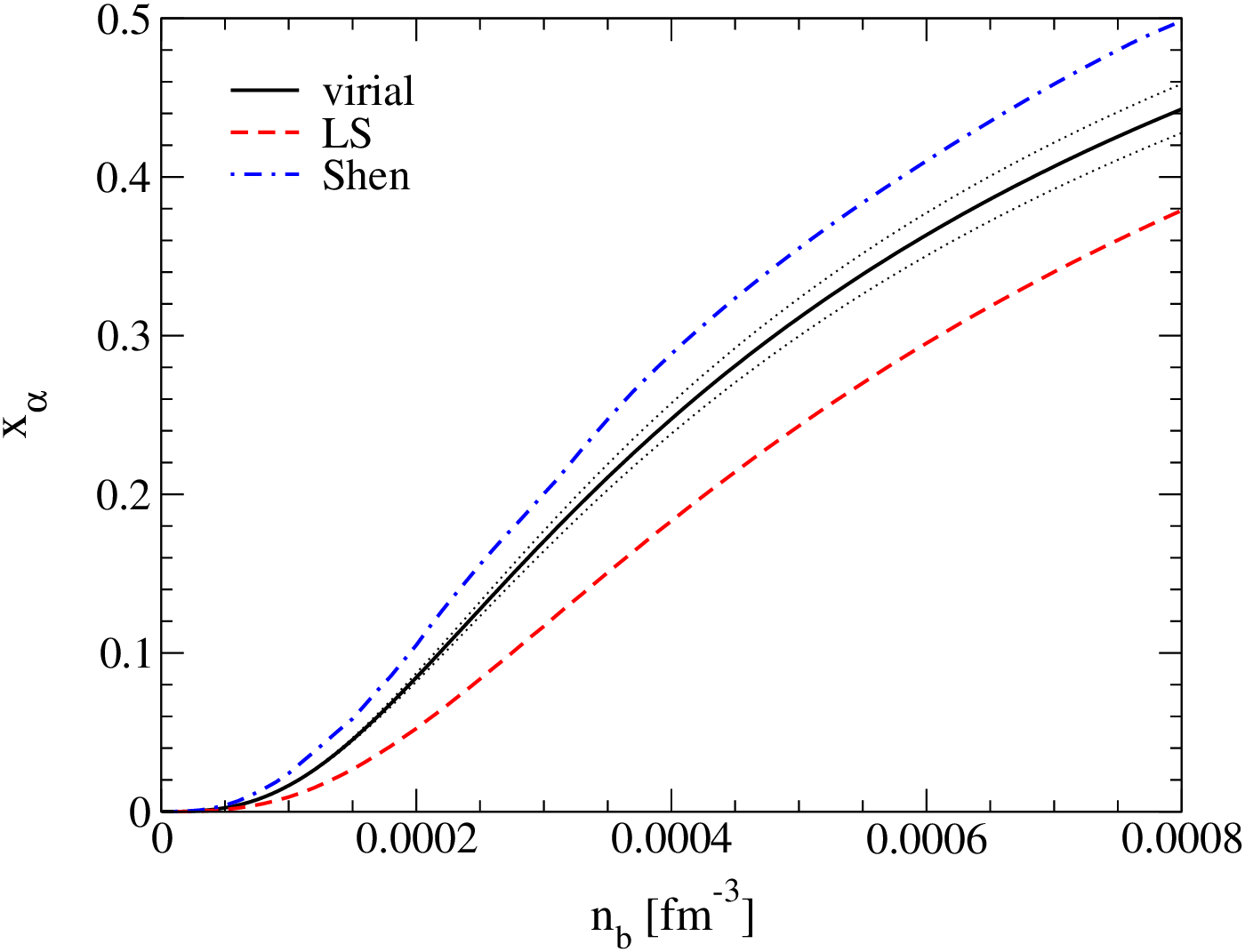}
\caption{(Color online) The $\alpha$ mass fraction $x_\alpha$ 
versus baryon density $n_b$ for $T=4 \mev$ and $Y_p=1/2$.  The 
dotted error band for the virial equation of state is based on 
an estimate of a neglected third virial coefficient 
$b_{\text{nuc}}^{(3)} = \pm 10$. Also shown are results for 
the LS~\cite{LS} and Shen~\cite{Shen} equations of state.}
\label{Fig5}
\end{center}
\end{figure}

The third virial coefficient $b_{\text{nuc}}^{(3)}$ can be used to
make a simple error estimate of neglected terms in the virial expansion. 
From the above conclusion, we expect $b_{\text{nuc}}^{(3)}$ to be 
the most important of the third virial coefficients, and we therefore
take into account only $b_{\text{nuc}}^{(3)}$ in this estimate.  
For simplicity, we consider symmetric nuclear matter with $z_p=z_n=z$,
for which the virial equation of state, Eqs.~(\ref{p},\ref{nn}-\ref{nalpha}),
including $b_{\text{nuc}}^{(3)}$ reads
\begin{align}
\frac{P}{T} &= \frac{4}{\lambda^3}(z+z^2b_{\text{nuc}}
+z^3b_{\text{nuc}}^{(3)}) \nonumber \\[1mm]
&+\frac{1}{\lambda_\alpha^3}(z_\alpha
+z_\alpha^2 b_\alpha +4z_\alpha z\,b_{\alpha n}) \,, \\[1mm]
n &= n_p+n_n =\frac{4}{\lambda^3}(z+2z^2 b_{\text{nuc}}
+3z^3b_{\text{nuc}}^{(3)}+8z_\alpha z\,b_{\alpha n}) \,, \\[1mm]
n_\alpha &= \frac{1}{\lambda_\alpha^3}(z_\alpha + 2z_\alpha^2 b_\alpha 
+ 4 z_\alpha z\,b_{\alpha n}) \,.
\end{align}
In Fig.~\ref{Fig5}, we again plot the $\alpha$ mass fraction.  The solid 
line is our previous result with $b_{\text{nuc}}^{(3)}=0$.  We also 
give approximate error bands by considering $b_{\text{nuc}}^{(3)}=10$ 
(which gives a lower $x_\alpha$) and $b_{\text{nuc}}^{(3)}=-10$.  
This estimates the error only from the effect of a 
typical neglected term in the virial expansion.  Our estimate 
$|b_{\text{nuc}}^{(3)}|<10$ is somewhat arbitrary, as it is based 
on the observation that all second virial coefficients are comfortably 
less then 10 for $T=4 \mev$.  We see from Fig.~\ref{Fig5} that the 
resulting error band is small.  {\it Therefore, our virial 
expansion makes a model-independent prediction for the $\alpha$ particle 
concentration.}  We also contrast our results in Fig.~\ref{Fig5} to
the $x_\alpha$ predicted by two phenomenological equations of state.  
The Lattimer-Swesty (LS) equation of state~\cite{LS} is based on an 
extended liquid drop model and is almost universally used in modern 
supernova simulations.  The LS model predicts too few alpha particles 
over the densities where the virial equation of state is applicable.  
Alternatively, the Shen-Toki-Oyamatsu-Sumiyoshi (Shen) equation of
state~\cite{Shen} is based on an approximate Thomas Fermi calculation
using a relativistic mean-field interaction. The Shen equation of state
predicts slightly too high values for $x_\alpha$ at this temperature.

\begin{figure}[t]
\begin{center}
\includegraphics[scale=0.36,clip=]{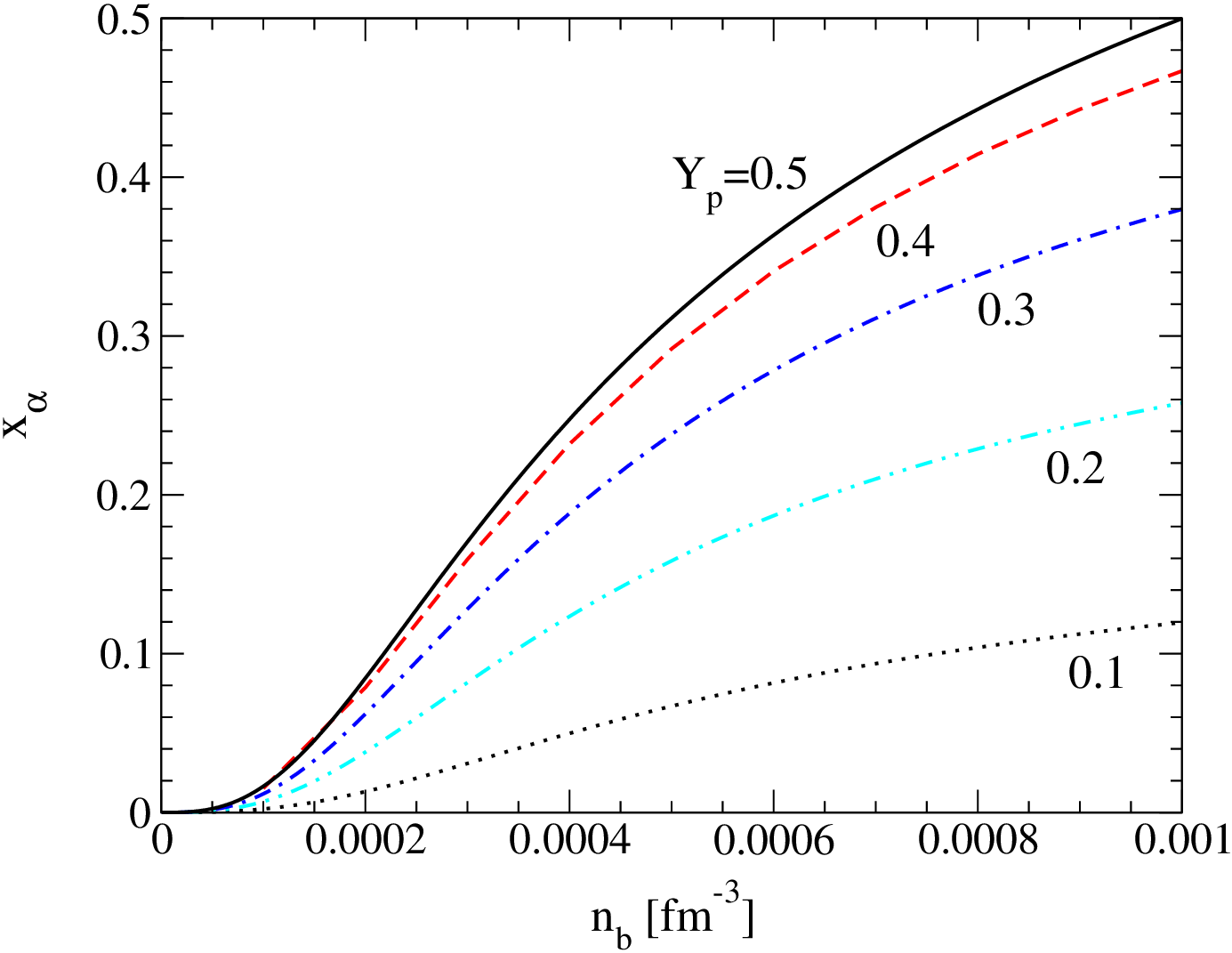}
\caption{(Color online) The $\alpha$ mass fraction $x_\alpha$ versus 
baryon density $n_b$ for $T=4 \mev$ and proton fractions $Y_p=0.1$, 
$0.2$, $0.3$, $0.4$ and $0.5$. For the lowest proton fraction $Y_p=0.1$,
the fugacities are $z_n < 0.2$, $z_p < 0.005$ and $z_\alpha < 0.001$ 
over this density range.}
\vspace*{-8mm}
\label{Fig6}
\end{center}
\end{figure}

\begin{figure}[t]
\begin{center}
\includegraphics[scale=0.36,clip=]{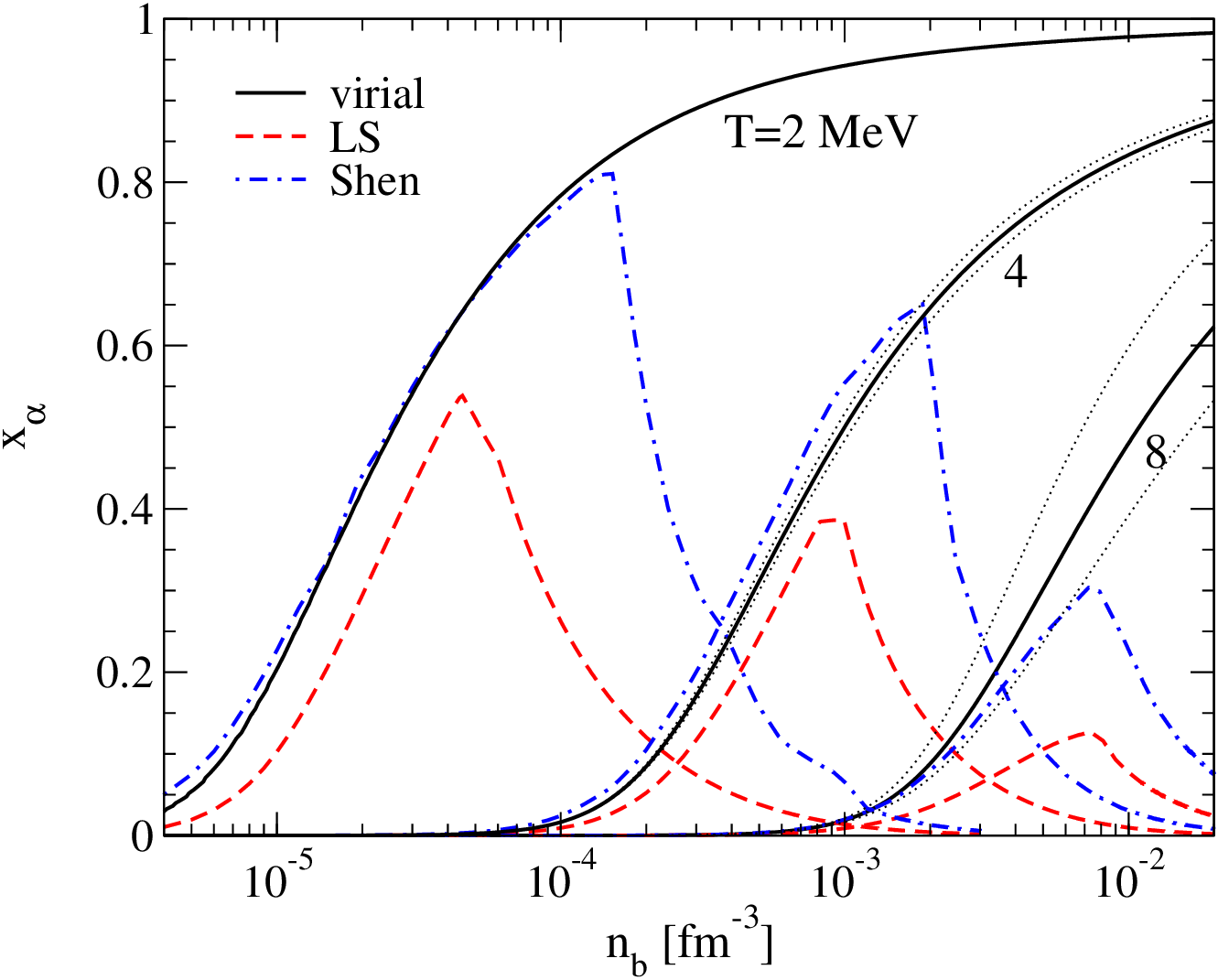}
\caption{(Color online) The $\alpha$ mass fraction $x_\alpha$ versus 
baryon density $n_b$ for $T=2$, $4$ and $8 \mev$ and $Y_p=1/2$.  
The dotted error bands for the virial equation of state are based on 
an estimate of a neglected third virial coefficient 
$b_{\text{nuc}}^{(3)} = \pm 10$. Note that for $T=2 \mev$ 
the error band is very small. Also shown are results for the 
LS~\cite{LS} and Shen~\cite{Shen} equations of state. Over this
density range, the fugacities are 
$z_n = z_p < 0.03$, $0.10$, $0.16$ and
$z_\alpha  < 0.27$, $0.09$, $0.03$ for $T=2$, $4$, $8 \mev$.}
\vspace*{-5mm}
\label{Fig7}
\end{center}
\end{figure}

\begin{figure}[t]
\begin{center}
\includegraphics[scale=0.36,clip=]{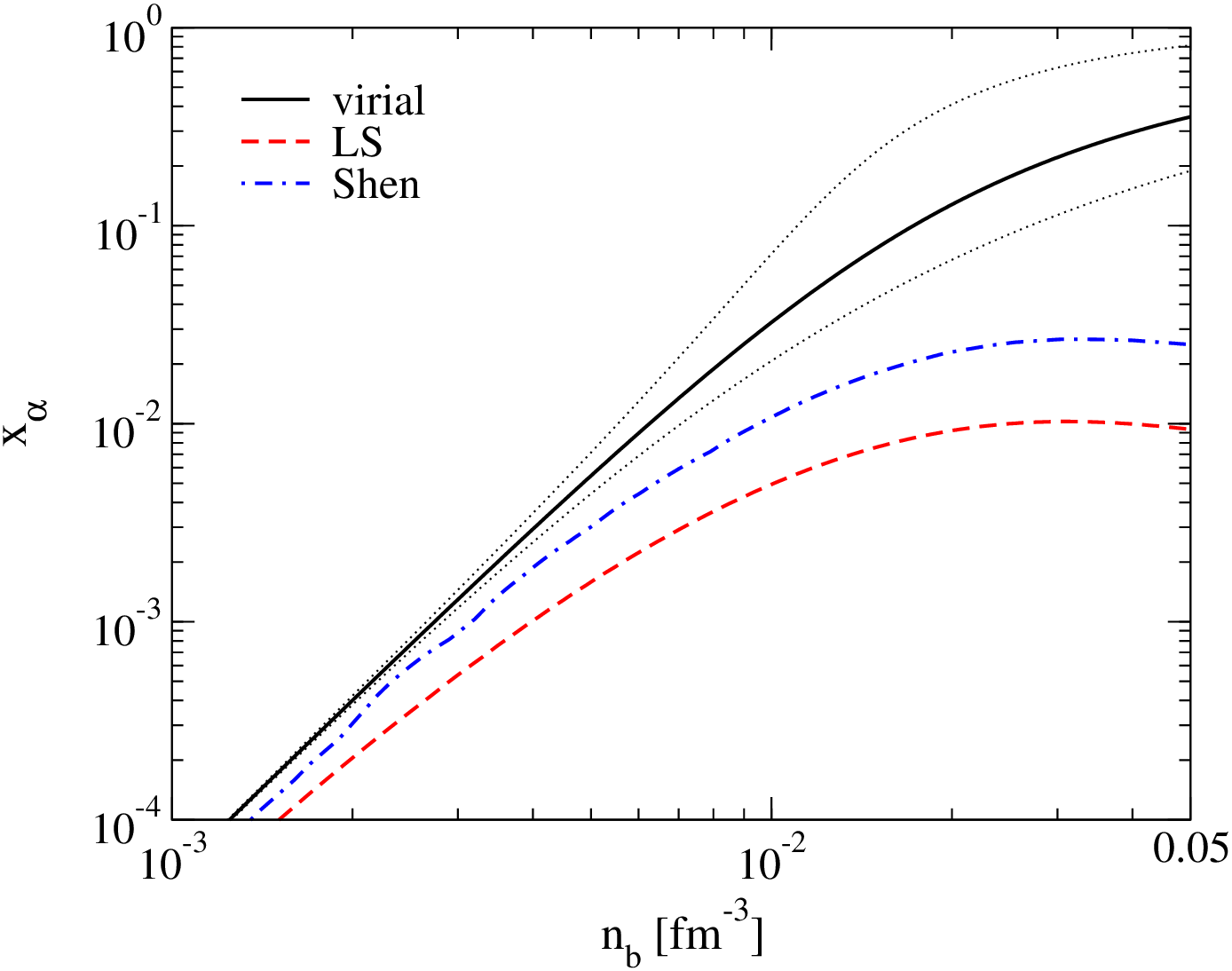}
\caption{(Color online) The $\alpha$ mass fraction $x_\alpha$ versus 
baryon density $n_b$ for $T=20 \mev$ and $Y_p=1/2$.  
The dotted error bands for the virial equation of state are based on 
an estimate of a neglected third virial coefficient 
$b_{\text{nuc}}^{(3)} = \pm 10$. Also shown are results for the 
LS~\cite{LS} and Shen~\cite{Shen} equations of state. Over this
density range, the fugacities are $z_n = z_p < 0.21$ and
$z_\alpha  < 0.07$.}
\vspace*{-7mm}
\label{Fig8}
\end{center}
\end{figure}

It is a simple matter to apply the virial equation of state to 
different proton fractions and temperatures. In Fig.~\ref{Fig6}, 
we show how the $\alpha$ particle concentration decreases with 
decreasing proton fraction $Y_p$ for $T=4 \mev$. In addition,
we give $x_\alpha$ for $Y_p=1/2$ and a range of 
temperatures $T=2$, $4$ and $8 \mev$ in Fig.~\ref{Fig7}.
At the lowest temperature $T=2 \mev$, $x_\alpha$ rises rapidly at 
low densities where the virial coefficients make only small 
contributions.  Therefore the error band is very small.  The error 
band becomes more important for $T=8 \mev$.  The LS equation of 
state is seen to significantly underestimate $x_\alpha$ 
at all three temperatures.  This may be due to a simple oversight 
in the published version~\cite{LS} of the LS equation of state.  Lattimer
and Swesty include the neutron-proton mass difference in the 
proton chemical potential, but do not appear to include twice 
this difference in the alpha chemical potential.  In contrast,
the Shen equation of state is close to our virial 
results at low densities. However, both LS and Shen equations
of state predict too few $\alpha$ particles at high temperatures 
$T > 10 \mev$. This can be clearly seen from 
the $\alpha$ mass fraction shown in Fig.~\ref{Fig8} for $T= 20 \mev$.
The differences of the LS and Shen equations of state at the higher
densities shown for $T= 20 \mev$ in Fig.~\ref{Fig8} are due to the
differences in the symmetry energy~\cite{Jim}.

\begin{figure}[t]
\begin{center}
\includegraphics[scale=0.36,clip=]{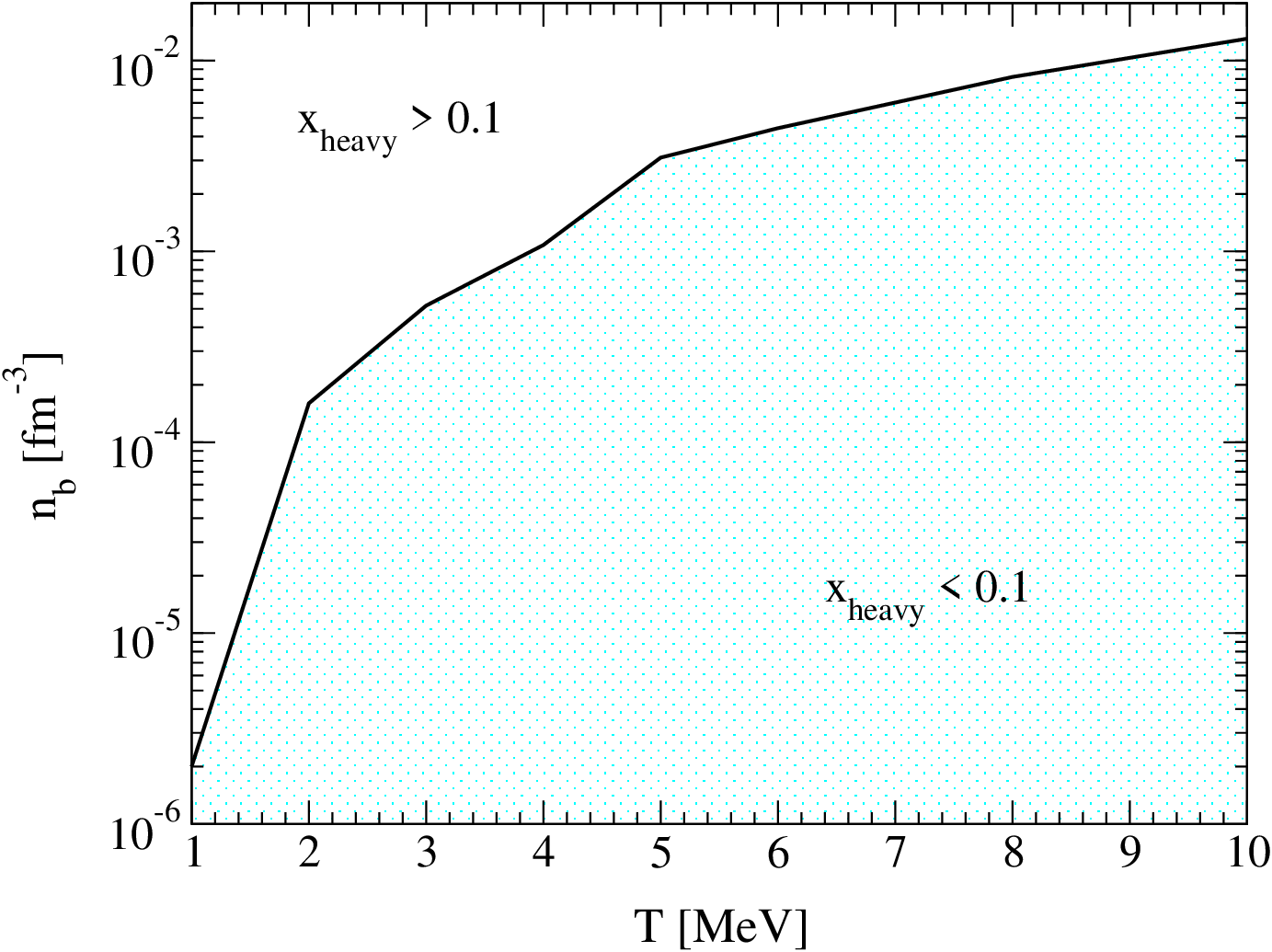}
\caption{(Color online) The threshold baryon density $n_b$ above 
which the Shen equation of state~\cite{Shen} predicts a heavy
nuclei mass fraction larger than 10\%. This figure is for $Y_p=1/2$.}
\label{Fig9}
\end{center}
\end{figure}

Both phenomenological equations of state shown in Fig.~\ref{Fig7} 
give $\alpha$ particle concentrations that first increase with density and 
then rapidly decrease.  This decrease is due to the formation of 
heavy nuclei. Heavy nuclei are not included in our first studies.
We will extend the virial formalism to include heavy nuclei in
future work. Therefore, our virial results are currently limited 
to lower densities, where one does not expect large contributions 
from heavy nuclei. In Fig.~\ref{Fig9}, we plot the density above 
which the Shen equation of state has a heavy nuclei mass fraction
larger than 10\%.  This provides an estimate for the range of validity 
of the virial $n$, $p$, $\alpha$ equation of state. For this 
estimate we have considered the Shen instead of the LS model due
to the LS error in $x_\alpha$.    

\begin{figure}[t]
\begin{center}
\includegraphics[scale=0.36,clip=]{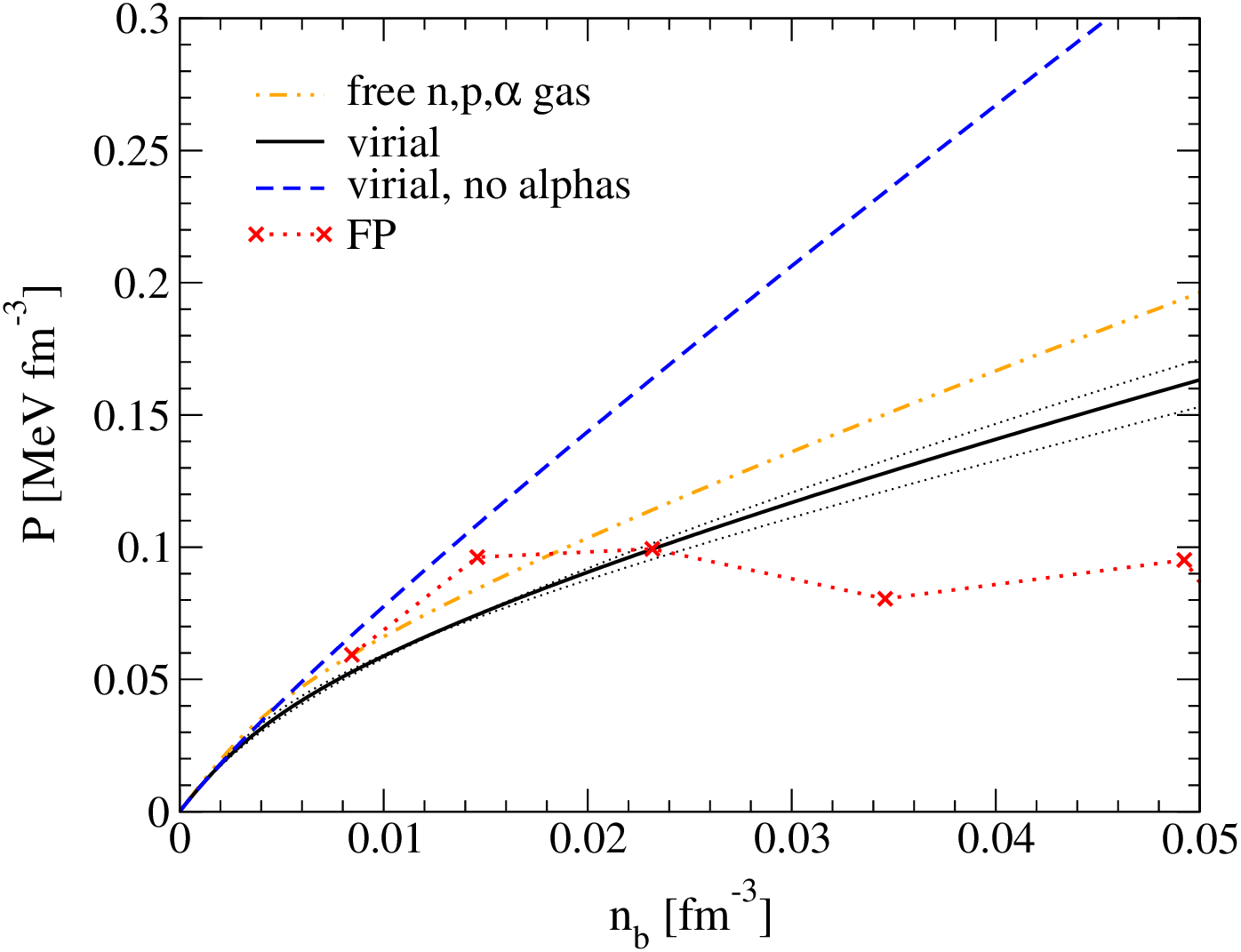}
\caption{(Color online) The pressure $P$ of symmetric nuclear 
matter versus baryon density $n_b$ for $T=10 \mev$. The dotted 
error band for the virial equation of state is based on an 
estimate of a neglected third virial coefficient 
$b_{\text{nuc}}^{(3)} = \pm 10$. Also shown are virial results 
without $\alpha$ particles ($z_\alpha=0$) and the microscopic 
FHNC results of Friedman and Pandharipande (FP)~\cite{FP}. 
The effects of interactions are illustrated by comparing to the
pressure of a free gas of nucleons and alpha particles.
Over this density range, the fugacities are $z_n = z_p < 0.21$ 
($< 0.51$ without alphas, $< 0.32$ for the free gas) and 
$z_\alpha  < 0.04$ ($< 0.17$ free gas).}
\label{Fig10}
\end{center}
\end{figure}

\subsection{Pressure}

Next, we present our results for the pressure.  In Fig.~\ref{Fig10}, 
we plot the pressure of symmetric nuclear matter for $T=10 \mev$.
The formation of $\alpha$ particles reduces the 
total number of particles in the system, and 
this significantly lowers the pressure. For comparison, we also show 
in Fig.~\ref{Fig10} the microscopic Fermi hyper-netted-chain (FHNC) 
equation of state of Friedman and Pandharipande~\cite{FP}.  The
latter calculation is based on a variational wave function with only
two-nucleon correlations.  However, the model-independent virial 
equation of state shows that there are large $\alpha$ concentrations 
in this density regime.  For example, at the threshold density of 
$n_b=0.013 \fm^{-3}$ (from Fig. \ref{Fig9}) we have $x_\alpha 
= 0.40$.  Microscopic calculations may require wave functions 
with four-nucleon correlations to accurately describe these $\alpha$ 
particle contributions.  {\it This suggests that all present microscopic 
variational or Brueckner equations of state are incomplete in their
description of low-density nuclear matter.}  In Fig.~\ref{Fig10},
we also give the pressure for densities beyond the threshold density
of Fig.~\ref{Fig9} in order to compare virial and FHNC results, which
both omit the formation of heavy nuclei.

The error band in Fig.~\ref{Fig10} is again estimated from
a neglected third virial coefficient $b_{\text{nuc}}^{(3)} = \pm 10$.  
For the pressure, this involves a cancellation. Increasing 
$b_{\text{nuc}}^{(3)}$ decreases the nucleonic contribution 
to the pressure due to the additional attractive interaction 
between nucleons. However, a larger $b_{\text{nuc}}^{(3)}$ also 
reduces the formation of $\alpha$ particles (due to a lower alpha
chemical potential), and this acts to increase the total pressure.           
Therefore, a positive $b_{\text{nuc}}^{(3)}$ reduces the pressure
at very low densities, but then leads to a larger pressure with
increasing density.

\begin{figure}[t]
\begin{center}
\includegraphics[scale=0.36,clip=]{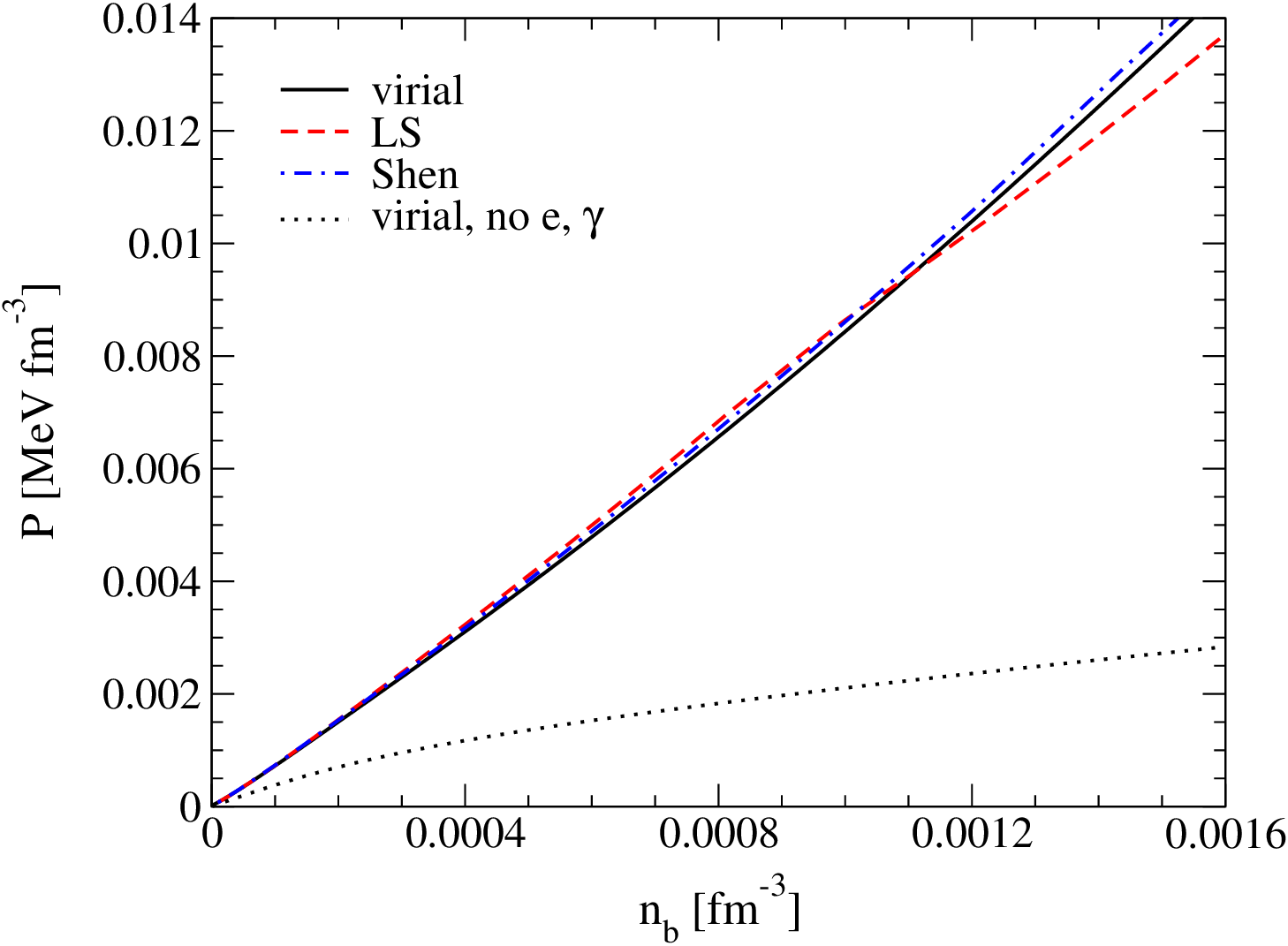}
\caption{(Color online) The pressure $P$ of symmetric nuclear 
matter versus baryon density $n_b$ for $T=4 \mev$. As discussed 
in the text, these results include the contribution of an electron
and photon gas, and we also show the virial pressure without this
contribution. In addition, we give the results of the LS~\cite{LS} 
and Shen~\cite{Shen} equations of state. Over this density range, 
the fugacities are $z_n = z_p < 0.06$ and $z_\alpha < 0.01$.}
\label{Fig11}
\end{center}
\end{figure}

In Fig.~\ref{Fig11}, we show our results for the pressure of
symmetric nuclear matter for $T=4 \mev$ in comparison to the
phenomenological LS and Shen equations 
of state.  The pressure given in Fig.~\ref{Fig11} also includes 
the contribution of an electron and photon gas as in Appendix~C
of~\cite{LS}. We find the pressures given by the virial equation 
of state and the phenomenological models agree well.

\begin{figure}[t]
\begin{center}
\includegraphics[scale=0.36,clip=]{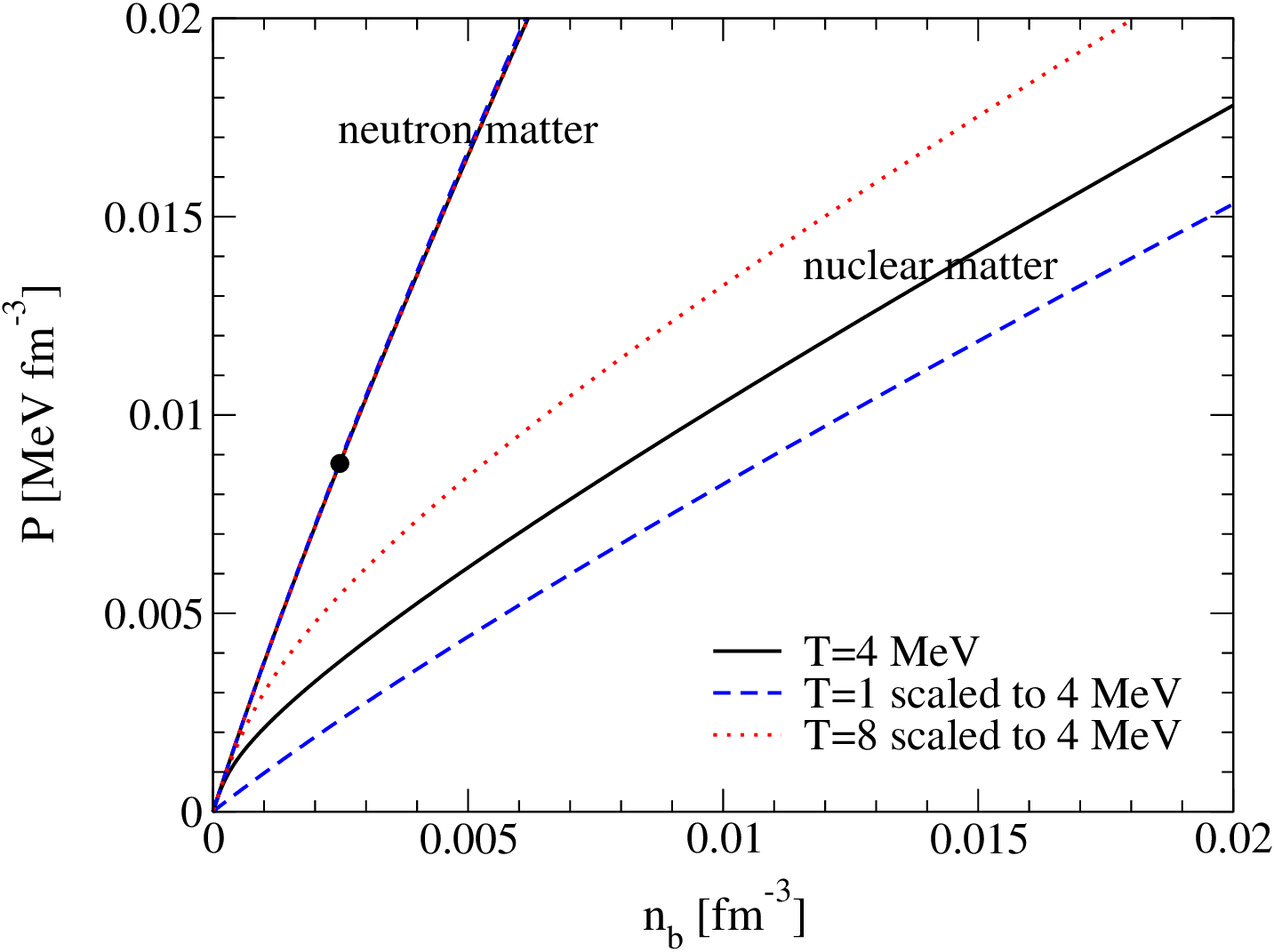}
\caption{(Color online) The pressure $P$ of pure neutron matter 
and of symmetric nuclear matter calculated for $T=1$, $4$ and 
$8 \mev$ and scaled to a temperature of $T'=4 \mev$ (for details 
see text). The circle indicates where $z_n =0.5$ for pure neutron
matter. For symmetric nuclear matter, the fugacities are 
$z_n = z_p < 0.10$ and $z_\alpha < 0.09$ over this density range.}
\label{Fig12}
\end{center}
\end{figure}

In a separate paper~\cite{vEOSneut}, we have demonstrated that the 
equation of state of low-density neutron matter scales to a
very good approximation.  Table~\ref{Table1} shows that the
neutron virial coefficient $b_n(T)$ is practically independent 
of the temperature.  If all virial coefficients are 
temperature independent, and provided the condition for
chemical equilibrium does not introduce a strong temperature dependence
through $z_\alpha(z_n,z_p,T)$, then the power series in the 
fugacity for the pressure and the baryon density will have no 
explicit temperature dependence.  In this case, the pressure 
will scale as
\begin{equation}
P(n_b,T) = T^{5/2} \, f(n_b/T^{3/2}) \,,
\end{equation}
where $f(x)$ is some function of $n_b/T^{3/2}$.  If this scaling
relation holds, one can predict the pressure $P(n_b,T')$ at a new 
temperature $T'$ from the original $P(n_b,T)$ through
\begin{equation}
P(n_b,T')=\biggl(\frac{T'}{T}\biggr)^{5/2} P(\widetilde{n}_b,T) \,,
\label{pscale}
\end{equation}
with $n_b = (T'/T)^{3/2} \, \widetilde{n}_b$.

We study this scaling symmetry in Fig.~\ref{Fig12}, 
where we have used Eq.~(\ref{pscale}) to predict 
the pressure of neutron matter for $T'=4 \mev$ from virial pressures 
calculated for $T=1$ and $8 \mev$.  The agreement with the unscaled
$T=4 \mev$ virial pressure is excellent.  This demonstrates that 
low-density neutron matter scales, even when interactions are included.  
Fig.~\ref{Fig12} also shows scaled pressures of symmetric nuclear matter 
for $T'=4 \mev$ using $T=1$ and $8 \mev$ results as input.  
These scaled pressures 
do not agree with the $T=4 \mev$ virial pressure.  Therefore, nuclear 
matter does not scale, which is due to clustering and mainly 
comes as a result of the strong 
temperature dependence of $z_\alpha$ and $x_\alpha$. 

\begin{figure}[t]
\begin{center}
\includegraphics[scale=0.36,clip=]{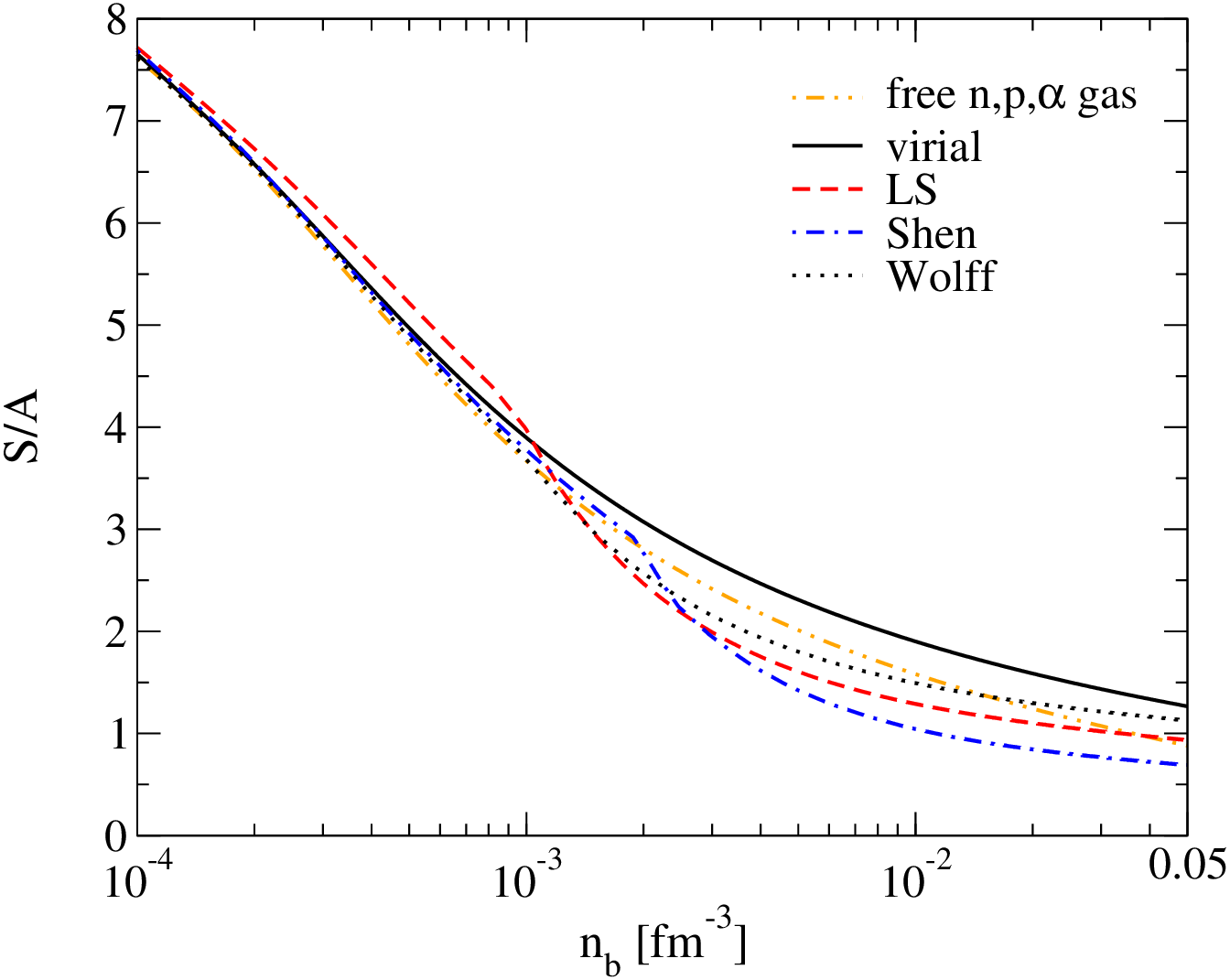}
\caption{(Color online) The entropy per baryon $S/A$ versus baryon 
density $n_b$ for $T=4 \mev$ and $Y_p=1/2$. As discussed in the text, 
these results include the contribution of an electron and photon gas. 
Also shown are results for the LS~\cite{LS}, Shen~\cite{Shen} and the 
Wolff~\cite{wolff} equations of state. 
The effects of interactions are illustrated by comparing to the
entropy of a free gas of nucleons and alpha particles.
Over this density range, 
the fugacities are $z_n = z_p < 0.11$ ($< 0.17$ for the free gas) 
and $z_\alpha < 0.17$ ($< 0.81$ free gas).}
\label{Fig13}
\end{center}
\end{figure}

\subsection{Entropy and Energy}

\begin{figure}[t]
\begin{center}
\includegraphics[scale=0.36,clip=]{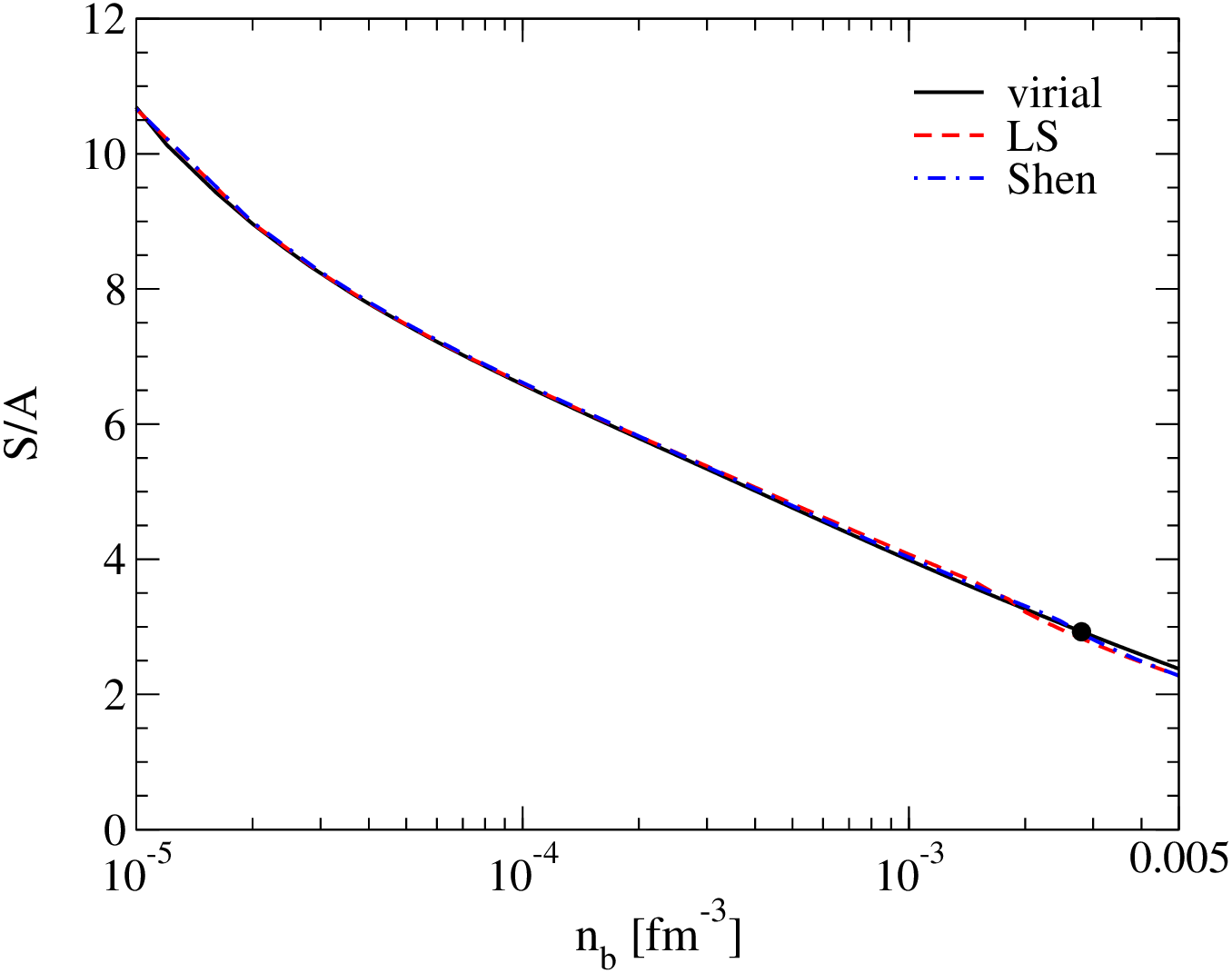}
\caption{(Color online) The entropy per baryon $S/A$ versus baryon 
density $n_b$ for $T=4 \mev$ and $Y_p=0.05$. As discussed in the text, 
these results include the contribution of an electron and photon gas. 
Also shown 
are results for the LS~\cite{LS} and the Shen~\cite{Shen} equations 
of state. The circle indicates where $z_n =0.5$ ($z_p = 0.002$ and 
$z_\alpha = 0.001$).}
\label{Fig14}
\end{center}
\end{figure}

The entropy per baryon $S/A$ of symmetric nuclear matter is shown 
for $T=4 \mev$ in Fig.~\ref{Fig13}.  Here, our virial calculation 
again includes the entropy of an electron and photon gas as in
Appendix~C of~\cite{LS}.  The entropy is clearly seen to 
reflect the composition.  At low densities, the LS entropy is 
above the other calculations because of the low $\alpha$ mass
fraction in the LS model.  At high densities, the entropy in the
phenomenological models is below our virial result due to the 
formation of heavy nuclei.  This deviation coincides with the threshold
density of Fig.~\ref{Fig9}.  In Fig.~\ref{Fig13}, we also show 
the entropy per baryon for the Wolff equation of state~\cite{wolff},
to give a measure for the model dependence of the onset 
of heavy nuclei. The entropy per baryon for extremely 
neutron-rich matter $Y_p=0.05$ is shown in Fig.~\ref{Fig14}.  Now 
there is very little change in composition, and thus good agreement 
between the virial and phenomenological results.

\begin{figure}[t]
\begin{center}
\includegraphics[scale=0.36,clip=]{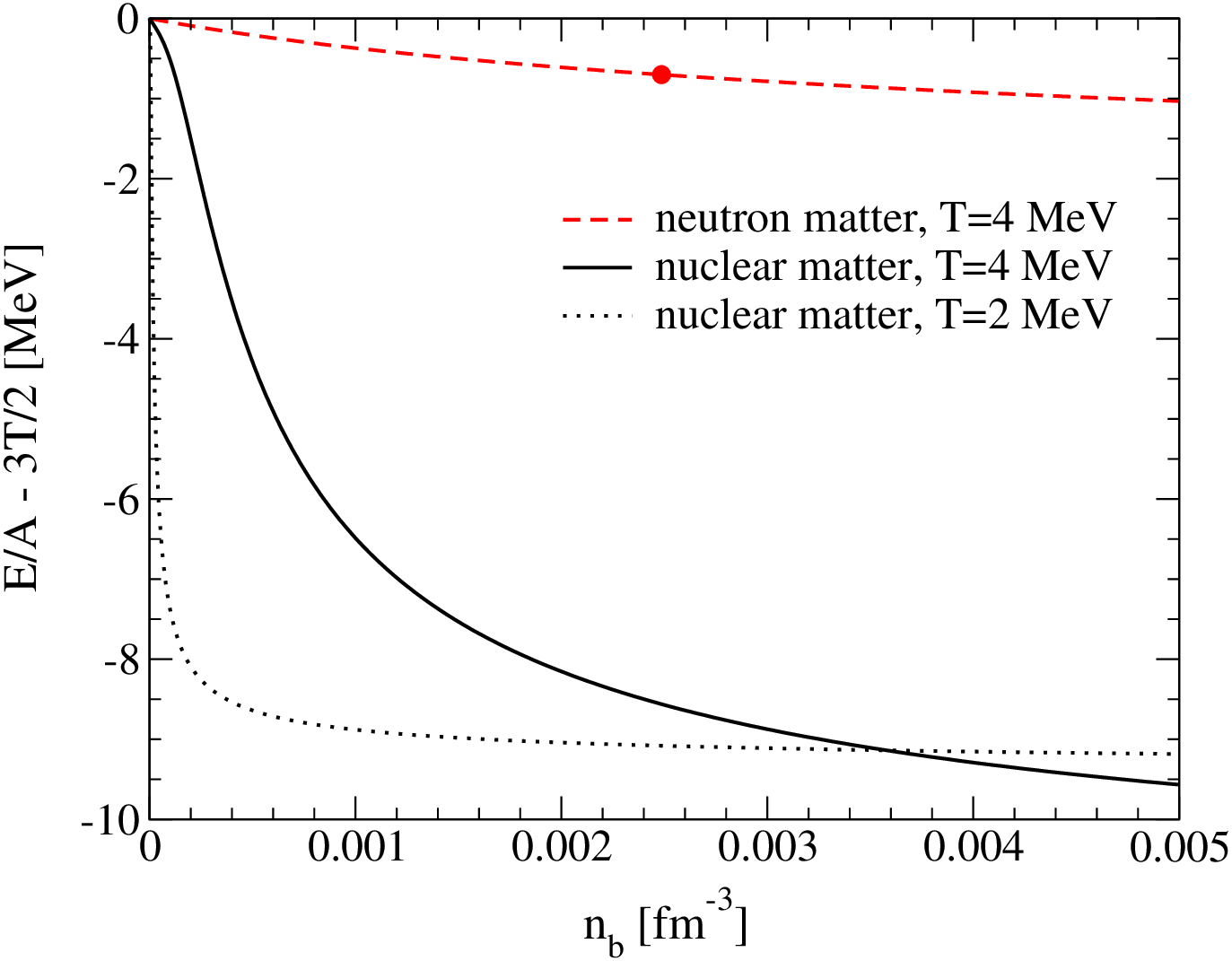}
\caption{(Color online) The energy per baryon $E/A$ versus baryon 
density $n_b$ for $T=4 \mev$ (and also $T=2 \mev$ (dotted) for 
nuclear matter). In order to clearly show the interaction 
effects we have subtracted the free kinetic energy $3T/2$ from 
all curves. The circle indicates where $z_n =0.5$ for pure neutron
matter. For symmetric nuclear matter, the fugacities are $z_n = z_p 
< 0.08$, $0.02$ and $z_\alpha < 0.03$, $0.11$ for $T=4$ and $2 \mev$ 
respectively.}
\label{Fig15}
\end{center}
\end{figure}

In addition to the entropy, we show our results for the energy
per baryon $E/A$ in Fig.~\ref{Fig15}.  In the limit of very low 
density, the energy reduces to the kinetic energy of an ideal 
classical gas $3T/2$.  Therefore, we have subtracted $3T/2$ from 
$E/A$ in Fig.~\ref{Fig15}, in order to clearly show the interaction 
effects. We find that the energy per baryon for pure neutron 
matter is somewhat below $3T/2$ due to the attractive nuclear 
interactions. Our neutron matter 
results are in excellent agreement with
the scaling law for the energy density $\epsilon = 3/2 P$,
which follows from Eq.~(\ref{epsilon}) when the virial coefficients 
are temperature independent, as is the case for neutron matter
with $b_n' \approx 0$. In contrast, $E/A$ for symmetric nuclear 
matter drops very rapidly with increasing density.  This reflects 
the large binding energy as $\alpha$ particles form. As can be seen
from Fig.~\ref{Fig15}, this drop becomes more rapid for lower 
temperatures.  We therefore expect that the inclusion of 
heavy nuclei, or many-nucleon correlations, are
necessary to obtain a constant energy per particle $E/A \approx 
-16 \mev$ for $T=0$ nuclear matter at subsaturation densities.

\subsection{Symmetry Energy}

\begin{figure}[t]
\begin{center}
\includegraphics[scale=0.36,clip=]{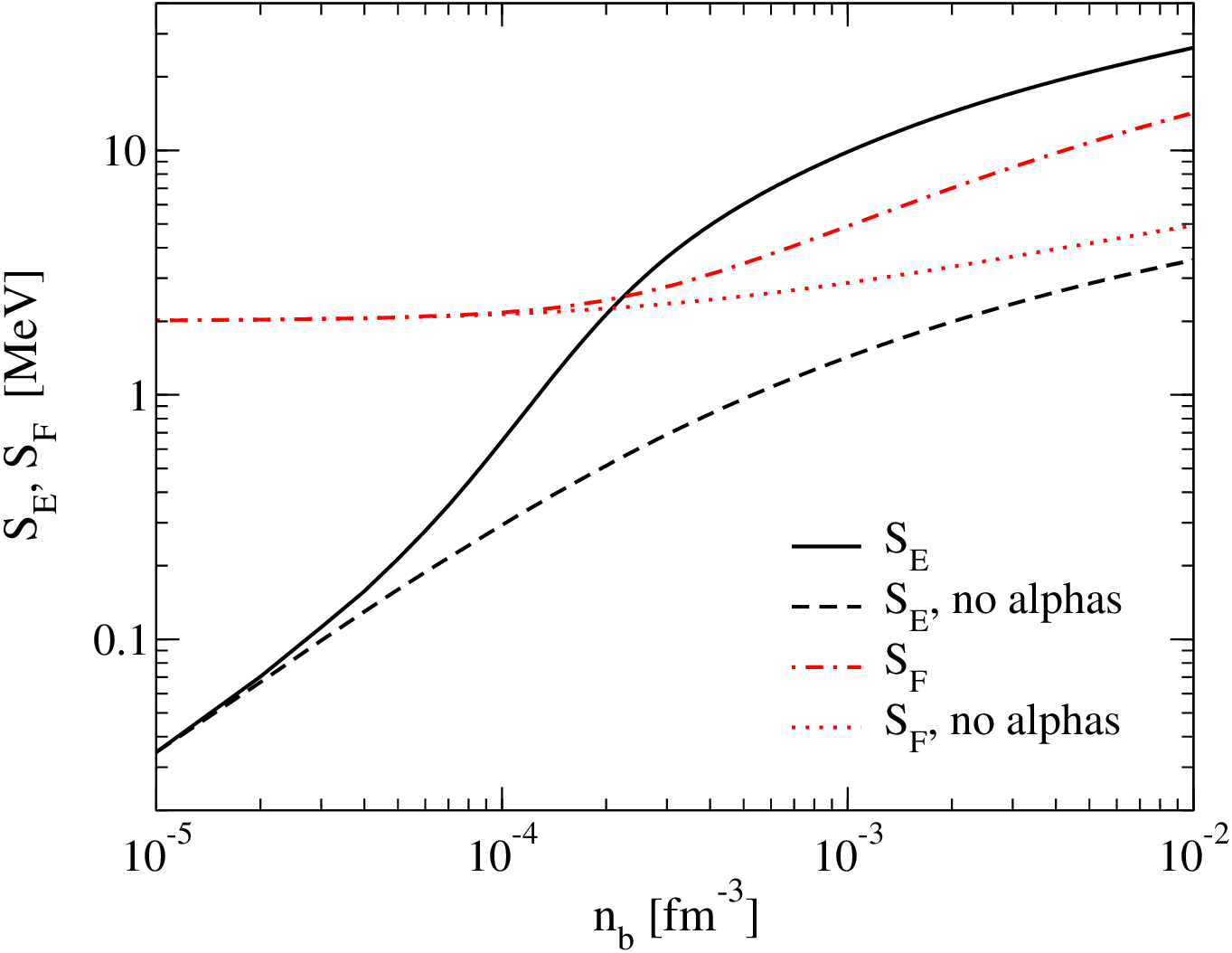}
\caption{(Color online) The symmetry energy $S_E$ and the symmetry 
free energy $S_F$ versus baryon density $n_b$ for $T=4 \mev$. 
Also shown are virial results without $\alpha$ particles 
($z_\alpha=0$). Over this density range, the fugacities are 
$z_n = z_p < 0.09$ ($< 0.37$ without alphas) and $z_\alpha  < 0.06$.}
\label{Fig16}
\end{center}
\end{figure}

The symmetry energy $S_E$ characterizes how the energy rises as one 
moves away from equal numbers of neutrons and protons,
\begin{equation}
S_E = \frac{1}{8} \frac{\partial^2}{{\partial Y_p}^2} 
\frac{E}{A} \: \biggr|_{Y_p=1/2} \,.
\label{se}
\end{equation}
We evaluate the second derivative of the energy per baryon with
respect to the proton fraction numerically. Fig.~\ref{Fig16}
shows the symmetry energy for $T=4 \mev$.  At very low density, 
$S_E$ rises slowly with density. To lowest order in the density
and neglecting $\alpha$ particles, the symmetry energy is given by
\begin{equation}
S_E \approx \frac{3}{8} \, T \lambda^3 n_b (b_{pn} - 2 T b_{pn}^\prime/3
- b_n + 2 T b_n^\prime/3) \,.
\end{equation}
Then, as $\alpha$ particles form, $S_E$ rises much faster with 
density.  As a result of clustering, the symmetry energy 
is large even at a very small fraction of saturation density. 
This is analogous to the observed behavior for the energy.

We also show in Fig.~\ref{Fig16} the symmetry free energy $S_F$ defined as
\begin{equation}
S_F = \frac{1}{8} \frac{\partial^2}{{\partial Y_p}^2} 
\frac{F}{A} \: \biggr|_{Y_p=1/2} \,.
\label{sf}
\end{equation}
To lowest order in 
the density and neglecting $\alpha$ particles, the symmetry free energy 
is given by
\begin{equation}
S_F \approx \frac{T}{2}+\frac{1}{4} \, T \lambda^3 n_b (b_{pn}-b_n) \,,
\end{equation}
where the $T/2$ term is from the entropy of mixing in the free 
energy density.  Again, as $\alpha$ particles form, we find in 
Fig.~\ref{Fig16} that $S_F$ increases rapidly with density.

\begin{figure}[t]
\begin{center}
\includegraphics[scale=0.36,clip=]{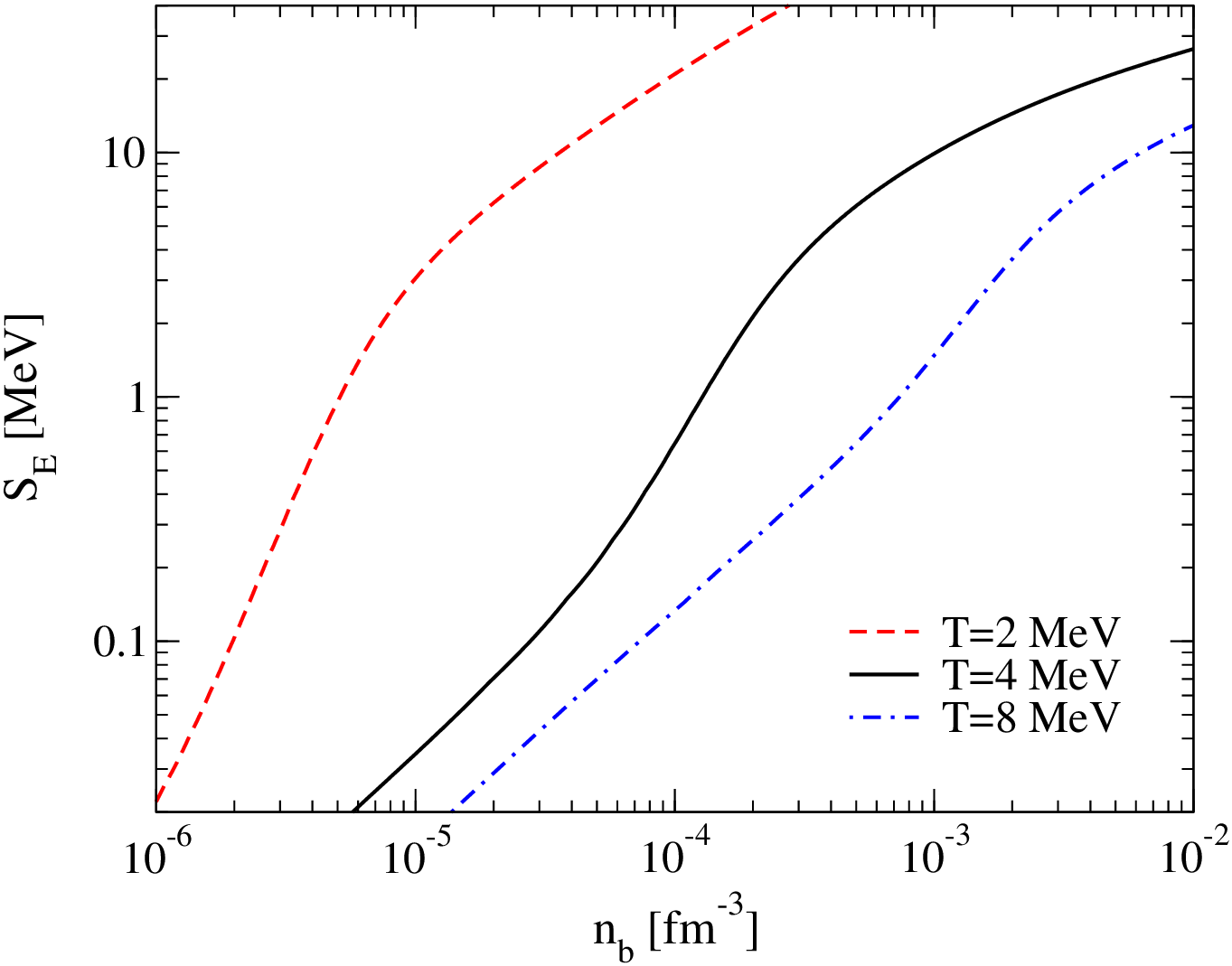}
\caption{(Color online) The symmetry energy $S_E$ versus baryon 
density $n_b$ for $T=2$, $4$ and $8 \mev$. Over this
density range, the fugacities are 
$z_n = z_p < 0.02$, $0.09$, $0.13$ and
$z_\alpha  < 0.18$, $0.06$, $0.01$ for $T=2$, $4$, $8 \mev$.}
\label{Fig17}
\end{center}
\end{figure}

Finally, the temperature dependence of the symmetry energy is shown 
in Fig.~\ref{Fig17} for $T=2$, $4$ and $8 \mev$.  The lower the 
temperature, the more rapidly $S_E$ rises with density. We conclude 
that, in the thermodynamic limit, the symmetry energy is large at
low densities due to cluster formation.

Isospin observables in asymmetric heavy-ion collisions, such as
the $N/Z$ ration of emitted fragments, are often analyzed with
semiclassical simulations, where nucleons move in isospin-dependent
mean fields and undergo two-nucleon collisions. The isospin-dependent
mean-field implies a certain symmetry energy. The density dependence
of this symmetry energy can be adjusted to improve agreement with
the data, e.g., see~\cite{tsang}. This phenomenological symmetry
energy at low densities is in general much smaller than our
virial $S_E$. This is because the symmetry energy in the virial 
expansion includes the contributions of clusters with high central
densities even at low baryon density. The relation between $S_E$
and the symmetry energy of mean-field models for single nuclei
should therefore be investigated in future work.

\section{Conclusions}
\label{conclusions}

We have presented the equation of state of low-density nuclear 
matter using the model-independent virial expansion. The virial 
equation of state is thermodynamically consistent and was 
calculated directly from binding energies
and scattering phase shifts, without the need of a nuclear
interaction model or nuclear wave functions. Therefore, our
results provide a benchmark for all microscopic or phenomenological
equations of state at low densities. The virial equation of state
can also be used to constrain models that extend to higher densities.
In some cases, our virial results may be used to constrain model 
parameters or functional forms for model interactions. In other cases, 
more extensive modifications of the phenomenological equations of state
may be necessary in order to reproduce the virial low-density limit.

Tightly bound nuclei can lead to unnaturally large bound state
contributions to the higher-order virial coefficients. Therefore,
we have considered a gas of nucleons and alpha particles to 
explicitly take into account the large $\alpha$ particle binding
energy. This leads to a larger radius of convergence compared to 
a nucleonic virial expansion. We have calculated the second
virial coefficients for NN, N$\alpha$ and $\alpha\alpha$ interactions
directly from the relevant binding energies and scattering phase
shifts. For the temperatures of interest, the virial coefficients
are dominated by the low-energy bound states and resonances, but
higher partial wave contributions are significant.

We have found that the second virial coefficient for neutron matter $b_n$
is approximately temperature-independent, which leads to an 
approximate scaling
symmetry of pure neutron matter. The second virial coefficients for
nuclear matter $b_{pn}$ and $b_{\text{nuc}}$ decrease with temperature,
and as a result nuclear matter does not scale. The second virial
coefficient for N$\alpha$ interactions $b_{\alpha n}$
is well approximated by the
resonant P$_{3/2}$ contribution. This holds over a wide range of
energies due to a cancellation among the additional partial waves.
The second virial coefficient for $\alpha\alpha$ interactions 
$b_\alpha$ increases with temperature, as
the $\alpha\alpha$ phase shifts become large at higher energies.
The inclusion of the Coulomb interaction for the virial coefficients, 
and the proper treatment
of inelasticities and mixing parameters (due to the tensor force)
are important topics for future improvements.

We have used the virial coefficients to make model-independent
predictions for a variety of properties of nuclear matter over
a range of densities, temperatures and composition. The
resulting $\alpha$ particle concentration disagrees with all
equations of state currently used in supernova simulations.
The contributions from low-energy resonances show most prominently 
in the composition of low-density nuclear matter.

The effects of the 
second virial coefficients follows a natural hierarchy, with the 
NN virial coefficients being more important than the N$\alpha$ 
virial coefficient, and the $\alpha\alpha$ virial coefficient
leading to very small changes. We have used this hierarchy
to make simple error estimates by studying the effects of a
neglected third virial coefficient $b_{\text{nuc}}^{(3)}=\pm 10$.
The resulting error bands are small. Detailed
investigations of the effects of omitted higher virial 
coefficients, as well as the effects of heavy nuclei will be
the topic of separate studies. For a better error estimate, it 
is important to also have a reliable calculation of the
third virial coefficients, which could, e.g., come from a
calculation in the effective field theory for halo nuclei~\cite{haloEFT}.
Furthermore, these error estimates can also help clarify the domain 
of validity of the virial expansion.

The physics of nuclear matter is very different from neutron
matter due to clustering. As the density increases, $\alpha$ 
particles form and this leads to a significant reduction of the 
pressure of low-density nuclear matter. Similarly, clustering 
increases the binding energy and it reduces the entropy, which 
reflects the number of particles in the system. As the density
increases further, heavier nuclei and larger clusters form, and
$\alpha$ particles become less important. In fact, the breakdown
of the virial expansion for nuclear matter is due to the formation
of heavy nuclei, which is reached before the nucleon fugacities 
become large. In nuclear matter at $T=0$, we expect that one 
has to explicitly include heavy nuclei, or many-nucleon correlations, 
to obtain a constant energy per baryon $E/A \approx -16 \mev$ and 
a constant symmetry energy at low densities. These virial results
greatly improve our conceptual understanding of nuclear matter at
subsaturation densities. 

The virial equation of state can be regarded as a systematic 
extension of NSE models for strong interactions in the resonant 
continuum. Such a comparison will also provide valuable insights
how to systematically organize the virial expansion to include
heavy nuclei. In addition, topics of future studies include
model-independent predictions for the neutrino response of
low-density nuclear matter and a detailed comparison with 
nuclear lattice calculations~\cite{Sfact}.

\section*{Acknowledgements}

We thank S. Karataglidis for the calculation of the nucleon-alpha 
phase shifts and A. Marek for providing us with tables of the
LS, Shen and Wolff equations of state.
This work is supported in part by the US DOE under Grant No. 
DE--FG02--87ER40365 and the NSF under Grant No. PHY--0244822.

\end{document}